\documentclass[oldversion]{aa} 
\usepackage{graphicx}
\usepackage{longtable}
\usepackage{txfonts}
\usepackage{rotating}
\usepackage{lscape}
\usepackage{color}
\newcommand{\gsim}{\;\lower.6ex\hbox{$\sim$}\kern-7.75pt\raise.65ex\hbox{$>$}\;}
\newcommand{\lsim}{\;\lower.6ex\hbox{$\sim$}\kern-7.75pt\raise.65ex\hbox{$<$}\;}

\begin{document}
\title{Chemical characterization of the globular cluster NGC~5634 
associated to the
Sagittarius dwarf spheroidal galaxy\thanks{Based on observations collected at 
ESO telescopes under programme 093.B-0583}\fnmsep\thanks{
   Tables 2 is only available in electronic form at the CDS via anonymous
   ftp to {\tt cdsarc.u-strasbg.fr} (130.79.128.5) or via
   {\tt http://cdsweb.u-strasbg.fr/cgi-bin/qcat?J/A+A/???/???}}
 }

\author{
E. Carretta\inst{1},
A. Bragaglia\inst{1},
S. Lucatello\inst{2},
V. D'Orazi\inst{2,3,4},
R.G. Gratton\inst{2},
P. Donati\inst{1,5},
A. Sollima\inst{1},
\and
C. Sneden\inst{6}
}

\authorrunning{E. Carretta et al.}
\titlerunning{Chemistry of NGC~5634}

\offprints{E. Carretta, eugenio.carretta@oabo.inaf.it}

\institute{
INAF-Osservatorio Astronomico di Bologna, Via Ranzani 1, I-40127 Bologna, Italy
\and
INAF-Osservatorio Astronomico di Padova, Vicolo dell'Osservatorio 5, I-35122
 Padova, Italy
\and
Department of Physics and Astronomy, Macquarie University, Sydney, NSW 2109, Australia
\and
Monash Centre for Astrophysics, School of Physics and Astronomy, Monash University, Melbourne, VIC 3800, Australia
\and
Dipartimento di Fisica e Astronomia, Universit\`a di Bologna, viale Berti Pichat
6, I-40127 Bologna, Italy
\and
Department of Astronomy and McDonald Observatory, The University of Texas,
Austin, TX 78712, USA
  }

\date{}

\abstract{As part of our on-going project on
the homogeneous chemical characterization of multiple stellar populations in 
globular clusters (GCs), we studied NGC~5634, associated to the
Sagittarius dwarf spheroidal galaxy, using high-resolution spectroscopy of red
giant stars collected with FLAMES@VLT. We present here the radial velocity 
distribution of the 45 observed stars, 43 of which are member, the detailed chemical abundance 
of 22 species for the seven stars observed with UVES-FLAMES, and the abundance of six elements for
stars observed with GIRAFFE. On our homogeneous UVES metallicity scale we 
derived a low metallicity [Fe/H]$=-1.867\pm0.019\pm0.065$ dex ($\pm$statistical
$\pm$systematic error) with $\sigma=0.050$ dex (7 stars). We found the normal
anti-correlations between light elements (Na and O, Mg and Al), signature of
multiple populations typical of  massive and old GCs. We confirm the associations
of NGC~5634 to the Sgr dSph, from which the cluster was lost a few Gyr ago, on the basis of
its velocity and position and the abundance ratios of
$\alpha$ and neutron capture elements.

}
\keywords{Stars: abundances -- Stars: atmospheres --
Stars: Population II -- Galaxy: globular clusters -- Galaxy: globular
clusters: individual: NGC~5634}
\maketitle

\section{Introduction}\label{intro}

Once considered good examples of simple stellar populations, Galactic globular
clusters (GCs) are currently thought to have formed in a complex chain of
events, which left a fossil record in their chemical composition  (see the
review by Gratton, Carretta \& Bragaglia 2012). Our homogeneous FLAMES survey 
of more than 25 GCs (see updated references in Carretta 2015 and Bragaglia et al.
2015)
combined with literature data, demonstrated that most, perhaps all, GCs host
multiple stellar populations that can be traced by the anticorrelated 
variations of Na and O abundances discovered by the Lick-Texas group (as
reviewed by Kraft 1994 and Sneden 2000). 
Photometrically, GCs exhibit spread, split and even multiple sequences, especially
when the right combination of filters are used.
These variations can be explained in large part by different chemical
composition among cluster stars, in particular of light elements like He, C, N,
O (e.g., Sbordone et al. 2011; Milone et al. 2012). 

Our large and homogeneous database allowed us for the first time a quantitative 
study of the Na-O anticorrelation. In all the analyzed GCs we found about one 
third of stars of primordial composition, similar to that of field stars of
similar metallicity (only showing a trace of type II Supernovae nucleosynthesis,
i.e. low  Na, high O). 
According to the most widely
accepted paradigm of GC formation (e.g., D'Ercole et al. 2008) these stars are
believed to be the long-lived part of the first generation
(FG) of  stars formed in the cluster. The other two thirds have a
modified composition (increased Na, depleted O) and belong to the second
generation (SG) of stars, polluted by the most massive stars of the FG (Gratton
et al. 2001) with ejecta from H burning at high temperature (Denisenkov \&
Denisenkova 1989, Langer et al. 1993). Unfortunately, what were the FG stars
that  produced the gas of modified composition is still an unsettled question,
see e.g.  Ventura et al. (2001),  Decressin et al. (2007), de Mink et al.
(2009), Maccarone \& Zureck (2012), Denissenkov \& Hartwick (2014), and Bastian
et al. (2015).

We found that the extension of the Na-O anticorrelation tends to be larger for
higher mass GCs and that, apparently, there is an observed minimum cluster mass
for appearance of the  Na-O anticorrelation (Carretta et al. 2010a).
This is
another important constraint for cluster formation mechanisms, because it
indicates the mass at which we expect that a cluster is able to retain part of
the ejecta of the FG, hence to show the Na-O signature (the masses of the
original clusters are expected to be much higher than the present ones, since
the SG has to be formed by the ejecta of only part of the FG). It is important
to understand if this limit is real or is due to the small statistics (fewer
low-mass clusters have been studied,  and only a few stars in each were
observed).  After studying the high-mass clusters, we begun a systematic study
of low-mass
GCs and high-mass and old open clusters (OCs) to empirically find the mass
limit for the appearance of the Na-O anticorrelation  and to understand if
there are differences between high-mass and low-mass cluster properties, e.g.
in the relative fraction of FG and SG stars (Bragaglia et al. 2012, 2014,
Carretta et al. 2014a).

For a better understanding of multiple stellar populations in GCs it is also 
fundamental to study clusters in other galaxies, a challenging task.  While a
promising approach seems to use abundance-sensitive colour indexes (see Larsen
et al. 2014 for GCs in Fornax), only a few GCs  (in Fornax and in the Magellanic
Clouds) have their
abundances derived using high-resolution spectroscopy\footnote{Of course
this does not include the known clusters associated with the Sgr dSph and now
physically within the Milky Way.}. These GCs also seem to
host two populations  (Letarte et al. 2006 for Fornax; Johnson et al. 2006 and
Mucciarelli et al. 2009 for old GCs in LMC) but the fractions of FG and SG stars
in Fornax and LMC GCs seem to be different with respect to clusters of 
similar mass in the Milky Way (MW).
Is this again a problem of low statistics or is the galactic environment (a
dwarf spheroidal and a dwarf irregular vs a large spiral) influencing the
GC formation mechanism?

To gain a deeper insight on this problem, we also included in our sample GCs 
commonly associated to the disrupting Sagittarius dwarf spheroidal to understand if there
is a significant difference amongst GCs formed in different environments (the MW
and dwarf galaxies). In fact, GCs born in a dSph may have retained a larger
fraction of their original mass. After the very massive GCs M54 (Carretta et al.
2010b) and NGC4590 (M~68: Carretta et al. 2009a,b; although this latter is not 
universally accepted as a member of the Sgr family), another Sgr GC of our
project is Terzan 8.  In this cluster we see some indication of a SG, at variance with
other low-mass  Sgr GCs (Ter7, Sbordone et al. 2007; Pal12, Cohen 2004).
However, the SG seems to represent a small
minority, contrary to what happens for  high-mass GCs (Carretta et al. 2014a).

In the present paper we focus on the chemical characterization of NGC~5634, a
poorly studied cluster considered to be associated to the Sgr dSph  (Bellazzini
et al. 2002, hereinafter B02). NGC~5634 is a relatively massive and metal-poor
GC ($M_V=-7.69$, [Fe/H]=-1.88; both values come from the 2010 web update
of the Galactic GC catalogue, Harris 1996).

The paper is organized as follows: in \S2 we present literature information on
the cluster, in \S3 we describe the photometric data, the spectroscopic
observations, and the derivation of atmosperic parameters. The abundance
analysis is presented in \S4,  a discussion on the light-element abundances
is given in \S5, the connection with Sgr dSph is discussed in \S6, and a summary
is presented in \S7.

\section{NGC~5634 in the literature}\label{lit}

The main studies on NGC~5634 were essentially focussed on verifying whether this
GC is associated to the Sagittarius galaxy, using either photometry (B02) or
spectroscopy (Sbordone at el. 2015, hereinafter S15). NGC~5634 is also part of
the study by Dias et al. (2016): they obtained FORS2@VLT spectra of  51 MW GCs
and determined metallicity and alpha elements (Mg, in particular) on a
homogeneous scale.

B02 observed this cluster with broadband $V,I$ Johnson filters, and using the
luminosity difference of the turnoff point with respect to the HB level they
concluded that NGC~5634 is as old as M~68 and Ter~8. The latter is 
still enclosed in the main body of Sgr and is considered one of the five
confirmed GCs belonging with high probability to this dwarf galaxy (see e.g.
Bellazzini et al. 2003, Law \& Majewski 2010a).
From the literature radial velocity and Galactocentric position B02 suggested
that NGC~5634 was a $former$ member of the Sgr galaxy that became unbound more
than 4 Gyrs ago. This deduction stems from the large distance (151 kpc) from the
main body of Sgr, a lag along the stream  that implies a rather large interval
since physical association.

Very recently S15 analyzed high resolution spectra taken with the HDS
spectrograph at the Subaru telescope (Noguchi et al. 2002) of two cool giants in NGC~5634,
obtaining the detailed abundances of about 20 species. They suggested  the
existence of multiple populations in the cluster from the anticorrelated
abundances of O, Na in the two stars, since the observed differences exceed any
spread due to the uncertainties associated to the abundance analysis.
At the low metallicity they derived for NGC~5634 (about [Fe/H]$=-1.98$ dex) the
overall chemical pattern of stars in dwarf galaxies is not so different from
that of the field stars of the Milky Way at the same metallicity, the largest
differences occurring at higher metallicity. Therefore, S15 were not able to
provide a clearcut chemical association of NGC~5634 to Sgr dSph, although they
concluded that an origin of this cluster in the Sgr system is  favoured by their
data.

Finally, Dias et al. (2016) analyzed spectra of nine stars in the wavelength
range  4560-5860 \AA, at a resolution about 2000. They measured  radial
velocities (RV) and found that eight of the stars are member of the cluster;
they also determined atmospheric parameters, iron and Mg (or alpha) abundance
ratios using a comparison with stellar libraries.  They found an average RV of 
about $-30$ km~s$^{-1}$, with a large dispersion (rms 39 km~s$^{-1}$), average
metallicity [Fe/H]=$-1.75$ dex (rms = 0.13 dex), average [Mg/Fe]=0.43 dex 
(rms= 0.02 dex), and average [$\alpha$/Fe]=0.20 dex (rms = 0.04) dex.  They did
not comment on anything peculiar for the cluster, they simply used it as part of
their homogenoeus sample.

\begin{table}
\centering
\setlength{\tabcolsep}{1.3mm}
\caption{Log of FLAMES observations.}
\begin{tabular}{lccccc}
\hline
Setup   &  UT Date   &  UT$_{init}$ & exptime & airmass & seeing\\
        & (yyyy-mm-dd) & (hh:mm:ss) & (s)	   & 	    & (arcsec) \\
\hline
HR11  & 2014-07-21 & 01:44:19.162 & 3600  & 1.302 &  0.74\\
HR11  & 2014-07-28 & 23:50:50.195 & 3600  & 1.096 &  0.77\\
HR11  & 2014-07-29 & 00:57:20.020 & 3600  & 1.244 &  0.65\\
HR11  & 2014-08-29 & 00:13:15.317 & 3600  & 1.581 &  0.71\\
\hline
\end{tabular}
\label{log}
\end{table}

\section{Observations and analysis}\label{obs}

\begin{figure}
\centering
\includegraphics[scale=0.5]{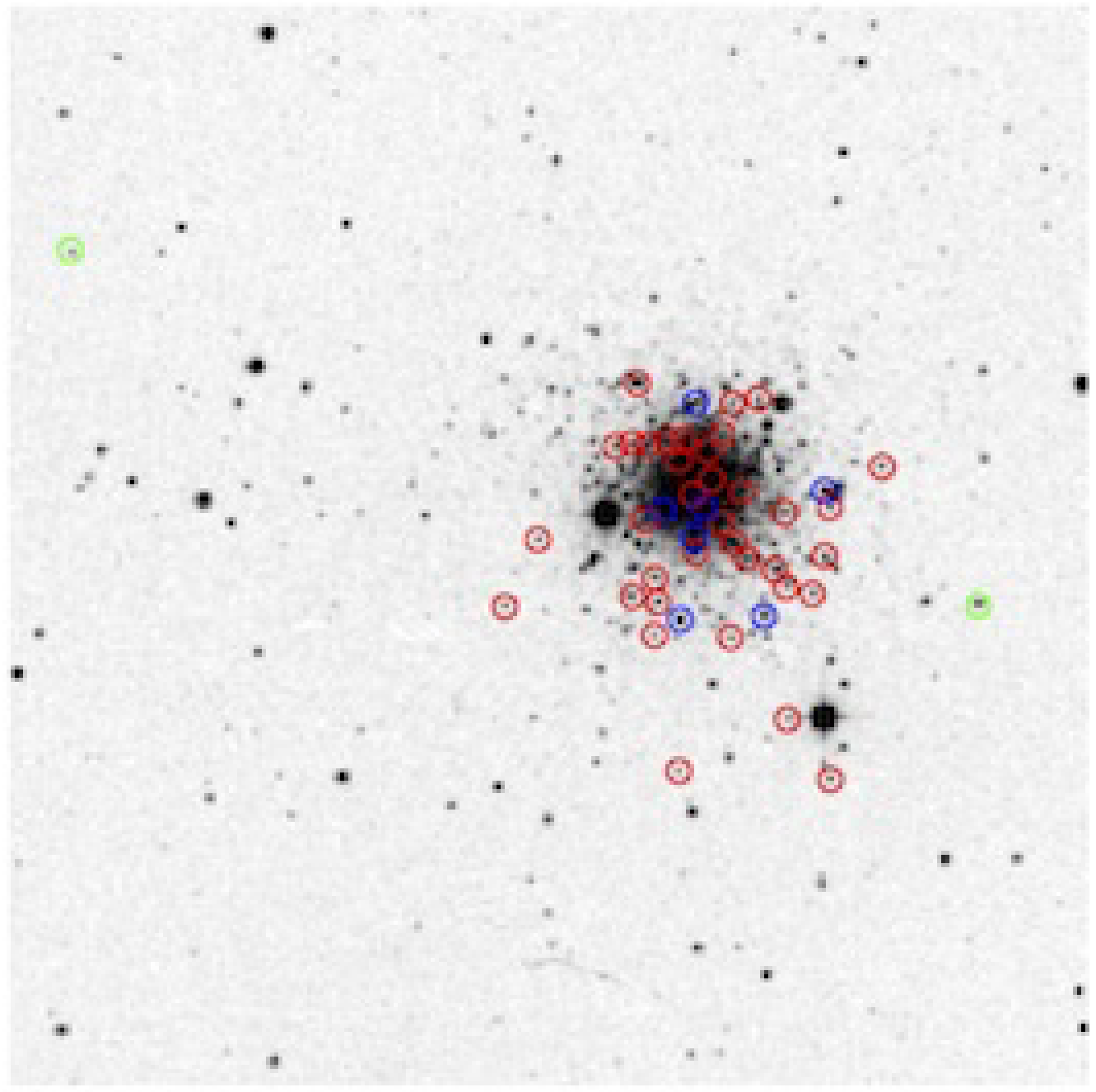}
\includegraphics[scale=0.43,bb=30 150 600 720,clip]{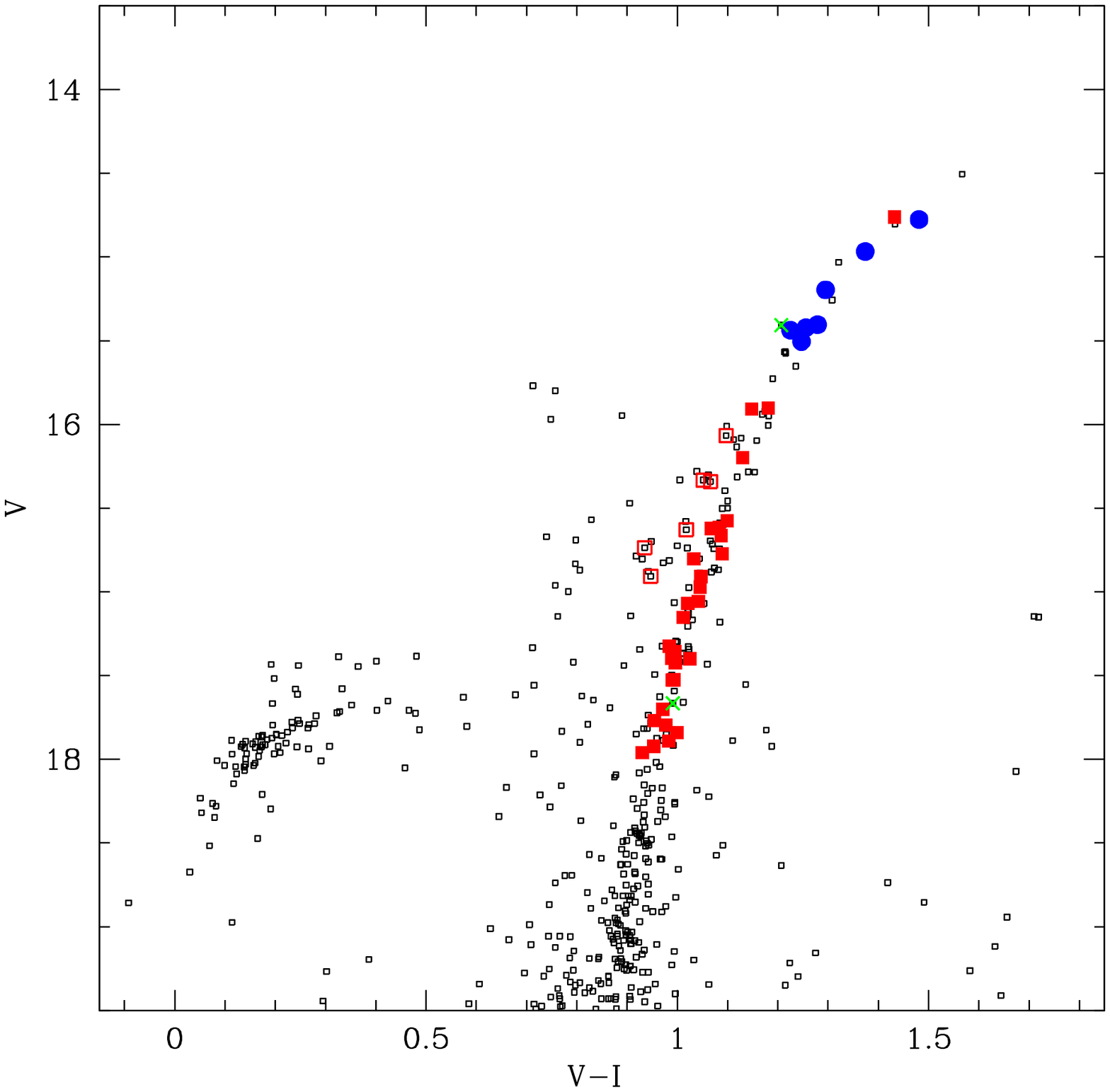}
\caption{Upper panel: a $15\arcmin \times 15\arcmin$ DSS map of NGC~5634, with
North up and East left. Our targets are colour-coded as observed with GIRAFFE
(in red), and UVES (in blue). Non member stars are indicated in green. Lower
panel: $V,V-I$ CMD (from B02) with FLAMES targets indicated by larger, coloured
symbols (subsample UVES: filled blue dots and subsample GIRAFFE filled and open
red squares for RGB and AGB stars, respectively.). Green crosses represent non
member stars.}
\label{cmdoss}
\end{figure}

We used the photometry by B02 to select our targets for FLAMES; the $V,V-I$\ CMD
is shown in Fig.~\ref{cmdoss}, lower panel.  We converted  the x,y positions
given in the catalogue to RA and Dec using stars in the {\em Two Micron All Sky
Survey} (2MASS, Skrutskie et al. 2006) for the astrometric
conversion.\footnote{We used the code {\sc cataxcorr}, developed by Paolo
Montegriffo at the INAF - Osservatorio Astronomico di Bologna, see
http://www.bo.astro.it/$\sim$paolo/Main/CataPack.htm }
We then selected stars on the red giant branch (RGB) and asymptotic giant branch
(AGB) and allocated targets using the ESO positioner {\sc fposs}. Given the
crowded field and the limitations of the instrument, only 45 targets were 
observed; they are indicated on a cluster map in Fig.~\ref{cmdoss}, upper
panel.

\subsection{FLAMES spectra}\label{spectra}

\begin{figure}
\centering
\includegraphics[scale=0.4,bb=50 150 600 720,clip]{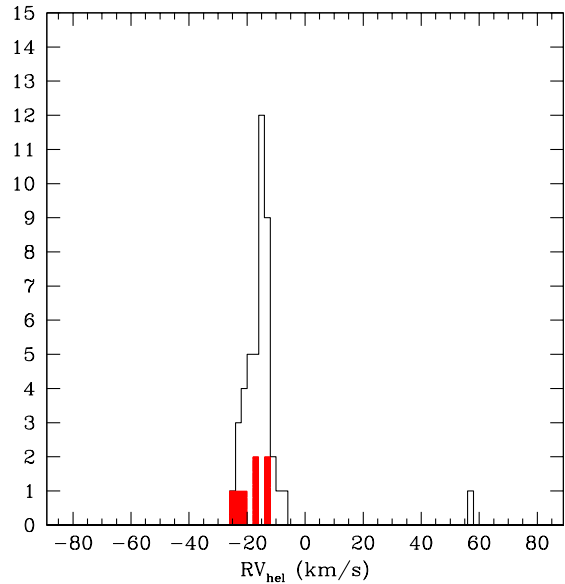}
\caption{Histogram of heliocentric RVs (the filled red histogram indicates the
seven UVES stars). The cluster stars are easily identified, with RV near $-16$
km~s$^{-1}$. }
\label{plotrv}
\end{figure}

NGC~5634 was observed with the multi-object  spectrograph FLAMES@VLT (Pasquini
et al. 2002) in the ESO program 093.B-0583 (PI A. Bragaglia).  The observations,
in priority B, were performed in service mode; a log is presented in
Table~\ref{log}.  Unfortunately, only less than half of the planned
observations  was actually completed. We only have four exposures (out of the
six requested) taken with the GIRAFFE high-resolution setup HR11 (R=24200), 
containing the
5682-88\AA \ Na {\sc i} doublet; no exposures with the HR13 setup,
containing the forbidden [O {\sc i}] lines,  are available. 
The GIRAFFE observations of 38 stars were coupled with the spectra of seven
stars obtained with the high-resolution (R=47000) UVES (Dekker et al.
2000) 580nm setup
($\lambda\lambda\simeq4800-6800$~\AA). Information on the 45 stars (ID,
coordinates, magnitudes and RVs) is given in Table~\ref{tabM}.

The reduced GIRAFFE spectra were obtained from the ESO archive (request 168411),
as part of the Advanced Data Products (ADP). The UVES spectra were reduced by us
using the ESO pipeline for UVES-FIBRE  data, which takes care of bias and flat
field correction, order tracing, extraction, fibre transmission, scattered
light, and wavelength calibration. We then used IRAF\footnote{IRAF is
distributed by the National Optical Astronomical Observatory, which are operated
by the Association of Universities for Research in Astronomy, under contract
with the National Science Foundation.} routines on the 1-d,
wavelength-calibrated individual spectra to subtract the (average) sky, measure
the heliocentric RV, shift to zero RV, and combine all the exposures for
each star.  

We show in Fig.~\ref{plotrv} the histogram of the RVs ; the cluster signature is
evident and we identitied 43 of the 45 observed targets as cluster members on
the basis of their RV.  One star has no measured RV, one  has a discrepant RV
and was labelled as non cluster member.  Star 151, initially considered member
of the cluster due to its RV, was afterward classified non-member following the
abundance analysis and then disregarded. The average, heliocentric RV for
each star is given in Table~\ref{tabM}, together with its rms.
For member stars we found an average RV of $-16.07$ km~s$^{-1}$ (with 
$\sigma=3.98$). This value is in excellent agreement with the average 
RV ($-16.7$ km~s$^{-1}$, $\sigma=5.5$) found by S15 from two stars, but not 
with the older literature value ($-45.1$ km~s$^{-1}$) listed in Harris (1996,
2010 web update), as already discussed in S15, or with the value of $-29.6$
($\sigma$=39.1) km~s$^{-1}$ in Dias et al. (2016), that was however obtained on much
lower resolution spectra.

\subsection{Atmospheric parameters}\label{paratmo}

We retrieved the 2MASS magnitudes (Skrutskie et al. 2006) of the 43 RV-member
stars; $K$ magnitudes, 2MASS identification, and quality flag are given in
Table~\ref{tabM}. Following our well tested procedure (for a detailed
description, see Carretta et al. 2009a,b),  effective temperatures $T_{\rm
eff}$\ were derived using an average relation between apparent magnitudes and
first-pass temperatures from $V-K$ colours and the calibrations of  Alonso et
al. (1999, 2001). This method allowed us to decrease the star-to-star errors in
abundances due to uncertainties in temperature.
The adopted reddening $E(B-V)=0.05$, distance modulus $(m-M)_V=17.16$, and input
metallicity [Fe/H]=$-1.88$ are taken from  Harris (1996, 2010 web update).
Gravities were obtained from apparent magnitudes and distance modulus, assuming
the  bolometric corrections from Alonso et al. (1999). We adopted a mass of
0.85~M$_\odot$\  for all stars and $M_{\rm bol,\odot} = 4.75$ as the bolometric
magnitude for the Sun, as in our previous studies.

Since only one of the two requested GIRAFFE setups was done, only a limited
number of Fe transitions from our homogeneous line list (from Gratton et al.
2003) were available for stars with GIRAFFE spectra, and this affected the
abundance analysis.
Fortunately, the seven stars with spectra taken with UVES were chosen among
those in the brightest magnitude range, close to the RGB tip  (see Fig.~\ref{cmdoss}) and
did not suffer excessively  for the lack of all planned observations. 
The S/N values per pixel range from 90 to 40 from the brightest down to the
faintest of our UVES sample stars. The median S/N for the  35 member stars with 
GIRAFFE spectra is 58 at $\lambda \sim 5600$ \AA. Literature high-resolution
spectra are available only for two stars in this cluster (S15), therefore even
such a small sample represents a significant improvement. 

\begin{figure}
\centering
\includegraphics[bb=19 162 555 613, clip,scale=0.42]{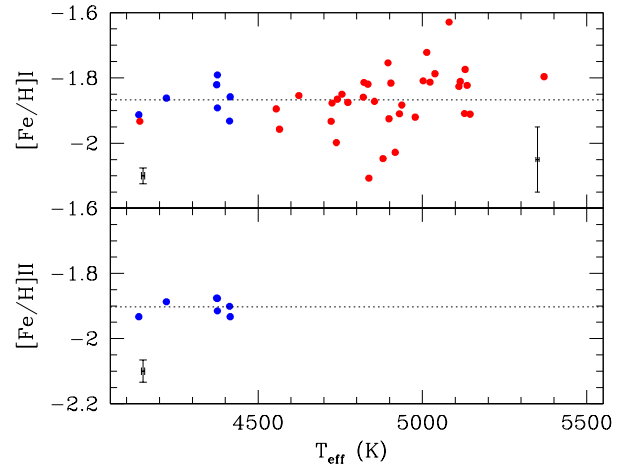}
\caption{Run of the iron abundances as a function of the effective temperature
for the seven UVES stars (blue circles) and stars with GIRAFFE spectra (red
circles). Abundances from singly ionized Fe lines are shown in the lower panel,
for UVES stars only. Internal error bars are also displayed (for the UVES sample
on the left corner, for the GIRAFFE sample in the right corner). The dotted line
is the average abundance derived from the UVES spectra.}
\label{feteffum34}
\end{figure}

We measured the equivalent widths (EW) of iron and other elements using the code
ROSA (Gratton 1988) as described in detail in Bragaglia et al. (2001).  We
employed spectrum synthesis for a few elements (see Sec.~\ref{neutron}). 
For the UVES spectra we
eliminated trends in the relation between abundances from Fe~{\sc i} lines and
expected  line strength (Magain 1984) to obtain values of the microturbulent
velocity $v_t$\footnote{For the 35 member stars with GIRAFFE HR11 spectra 
too few Fe lines were available and we generally adopted a relation with
the star's gravity: $v_t=-0.31 \log g +2.19$, apart from a few 
cases, derived from our previous analyses.}. Finally we interpolated  within the
Kurucz (1993) grid of model atmospheres (with overshooting on) to derive the
final abundances, adopting for each star the model with the appropriate
atmospheric parameters and whose abundances matched those derived from  
Fe {\sc i} lines.
The adopted atmospheric parameters ($T_{\rm eff}$, $\log g$, [A/H], and $v_t$)
are listed in Table~\ref{tabatmo} together with iron abundances. 
Figure~\ref{feteffum34} shows the run of [Fe/H] from
neutral and ionised transitions as a function of the effective temperature; no 
trend is visible in either case.
We find for NGC~5634 an average metallicity [Fe/H]$=-1.867 \pm 0.019$ dex
(rms=0.050 dex) and  $-1.903 \pm 0.009$ dex (rms=0.025 dex) from the seven stars
with UVES spectra, respectively for  neutral and ionised lines. The average
difference $-0.036 \pm 0.015$ dex (rms=0.040) dex is not significant.

The average value is in very good agreement with the mean metallicity we derived
from the analysis of the 35 stars with GIRAFFE spectra: 
[Fe/H]$=-1.869 \pm 0.016$ dex ($\sigma=0.093$ dex). The larger dispersion is
ultimately mostly due to the  limited number of Fe lines available  in the
spectral range of the HR11 setup, hampering a better derivation of the
parameters (see also Section \S4).

\setcounter{table}{2}
\begin{table*}
\centering
\caption[]{Adopted atmospheric parameters and derived metallicity.}
\begin{tabular}{rccccrcccrccc}
\hline
ID   &   $T_{\rm eff}$ &  $\log$ $g$ &  [A/H]  & $v_t$	     &  nr &  [Fe/H]{\sc i} &  rms &  nr &  [Fe/H{\sc ii} &  rms  \\
       &      (K)	&   (dex)     &  (dex)  & (km s$^{-1}$) &     &  (dex) 	  & 	  &     &  (dex)               \\
\hline
\multicolumn{11}{c}{UVES sample} \\
 2 &4137 &0.61 &-1.91 &2.11 &80 &-1.913 &0.126  &15 &-1.933 &0.077 \\ 
 5 &4221 &0.76 &-1.86 &2.10 &73 &-1.862 &0.095  &12 &-1.887 &0.100 \\ 
 7 &4376 &1.06 &-1.89 &1.84 &73 &-1.892 &0.130  &18 &-1.915 &0.104 \\ 
 9 &4376 &1.04 &-1.79 &2.00 &78 &-1.791 &0.101  &14 &-1.877 &0.069 \\ 
10 &4374 &1.04 &-1.82 &1.88 &73 &-1.821 &0.110  &12 &-1.876 &0.075 \\ 
12 &4413 &1.12 &-1.93 &1.66 &55 &-1.932 &0.112  &14 &-1.901 &0.087 \\ 
13 &4415 &1.11 &-1.86 &1.91 &60 &-1.858 &0.167  & 6 &-1.933 &0.090 \\ 

\multicolumn{11}{c}{GIRAFFE sample} \\

100&4854 &1.89 &-1.87 &1.60 &11 &-1.872 &0.184  &   &       &      \\ 
103&4904 &2.01 &-1.82 &1.57 & 8 &-1.816 &0.163	&   &	    &	   \\ 
113&4898 &1.98 &-1.93 &1.38 & 8 &-1.925 &0.131	&   &	    &	   \\ 
125&4930 &2.06 &-1.91 &1.55 & 5 &-1.910 &0.041	&   &	    &	   \\ 
132&5023 &2.24 &-1.81 &1.50 & 5 &-1.813 &0.172	&   &	    &	   \\ 
135&5013 &2.21 &-1.71 &1.50 & 4 &-1.722 &0.091	&   &	    &	   \\ 
142&5002 &2.18 &-1.81 &1.90 & 6 &-1.809 &0.135	&   &	    &	   \\ 
146&5038 &2.26 &-1.79 &1.49 & 3 &-1.787 &0.259	&   &	    &	   \\ 
152&4978 &2.14 &-1.92 &1.53 & 5 &-1.920 &0.151	&   &	    &	   \\ 
163&5081 &2.33 &-1.63 &1.47 & 5 &-1.629 &0.047	&   &	    &	   \\ 
167&5111 &2.38 &-1.83 &1.45 & 4 &-1.826 &0.127	&   &	    &	   \\ 
169&5130 &2.42 &-1.77 &1.81 & 4 &-1.774 &0.048	&   &	    &	   \\ 
173&5115 &2.39 &-1.81 &1.76 & 5 &-1.811 &0.142	&   &	    &	   \\ 
182&5128 &2.41 &-1.91 &1.44 & 3 &-1.909 &0.407	&   &	    &	   \\ 
189&5136 &2.43 &-1.82 &1.44 & 4 &-1.823 &0.120	&   &	    &	   \\ 
194&5145 &2.45 &-1.91 &1.43 & 4 &-1.911 &0.129	&   &	    &	   \\ 
19 &4555 &1.37 &-1.90 &1.77 & 8 &-1.895 &0.098	&   &	    &	   \\ 
22 &4565 &1.39 &-1.96 &1.76 & 7 &-1.957 &0.103	&   &	    &	   \\ 
30 &4820 &1.62 &-1.86 &1.69 & 5 &-1.859 &0.204	&   &	    &	   \\ 
35 &4624 &1.49 &-1.85 &1.73 & 9 &-1.854 &0.145	&   &	    &	   \\ 
3  &4140 &0.62 &-1.93 &2.00 &10 &-1.933 &0.098	&   &	    &	   \\ 
45 &4725 &1.68 &-1.88 &1.67 & 6 &-1.877 &0.115	&   &	    &	   \\ 
46 &4722 &1.67 &-1.93 &1.67 & 8 &-1.933 &0.207	&   &	    &	   \\ 
54 &4741 &1.70 &-1.87 &1.66 & 6 &-1.865 &0.119	&   &	    &	   \\ 
58 &4755 &1.73 &-1.85 &1.65 & 6 &-1.850 &0.099	&   &	    &	   \\ 
60 &4738 &1.69 &-2.00 &1.67 & 4 &-1.998 &0.110	&   &	    &	   \\ 
62 &4773 &1.76 &-1.87 &1.64 & 6 &-1.875 &0.147	&   &	    &	   \\ 
64 &4937 &1.90 &-1.88 &1.60 & 7 &-1.883 &0.177	&   &	    &	   \\ 
70 &4834 &1.89 &-1.82 &1.60 & 5 &-1.819 &0.128	&   &	    &	   \\ 
72 &4880 &1.96 &-2.05 &1.23 & 4 &-2.047 &0.018	&   &	    &	   \\ 
78 &4837 &1.88 &-2.11 &0.37 & 6 &-2.107 &0.188	&   &	    &	   \\ 
82 &4917 &1.93 &-2.03 &0.83 & 3 &-2.028 &0.018	&   &	    &	   \\ 
87 &4822 &1.84 &-1.81 &1.62 & 6 &-1.814 &0.157	&   &	    &	   \\ 
93 &4895 &2.00 &-1.88 &1.57 & 5 &-1.754 &0.132	&   &	    &	   \\ 
97 &5370 &2.21 &-1.80 &1.50 & 4 &-1.796 &0.117	&   &	    &	   \\ 
\hline
\end{tabular}
\label{tabatmo}
\end{table*}

\section{Abundances}\label{abu}
Beside Fe, we present here abundances of O, Na, Mg, Al, Si, Ca, Sc, 
Ti (both from neutral and singly ionized transitions), V, Cr (from both 
Cr~{\sc i} and Cr~{\sc ii} lines), Mn, Co, Ni, Zn, Cu, Y,
Zr, Ba,  La, Ce, Nd, and Eu, obtained from UVES spectra.
For stars in the GIRAFFE sample we derived only abundances of  one proton-capture
element (Na), three $\alpha-$capture elements (Mg, Si, and Ca), and three
elements of the iron-group (Sc, V, and Ni).
The abundances were derived using $EW$s for all species except Cu and
neutron-capture elements. The atomic data for the lines  and the solar reference
values come from Gratton et al. (2003).  The Na abundances were corrected for
departure from local thermodynamical equilibrium according to Gratton et al.
(1999), as in all the other papers of our FLAMES survey.
Corrections to account for the hyperfine structure were applied to Sc, V, and
Mn (references are in Gratton et al. 2003), and Y.

To estimate the error budget we closely followed the procedure described in
Carretta et al. (2009a,b). Table~\ref{sensu} provides the sensitivities of
abundance ratios to uncertainties in atmospheric parameters and $EW$s and the 
internal and systematic errors relative to the abundances from UVES spectra.
The same quantities are listed in Table~\ref{sensm} for abundances from GIRAFFE
spectra. In this second case, to have a conservative estimate of the internal 
error in $v_t$ we adopted the quadratic sum of the GIRAFFE internal errors for
the metal-poor GCs in our FLAMES survey: NGC~4590, NGC~6397, NGC~6805, NGC~7078,
NGC~7099 (Carretta et al. 2009a), NGC~6093 (Carretta et al. 2015), and NGC~4833
(Carretta et al. 2014b).

The sensitivities were obtained by repeating the abundance analysis
for all stars,  while changing one atmospheric parameter at the time, then
taking the average. The amount of the variation in the input parameters used in
the sensitivity computations is given in the table headers. 

On our UVES metallicity scale (see Carretta et al. 2009c) the average metal
abundance for NGC~5634 is therefore 
[Fe/H]$=-1.867\pm 0.019\pm 0.065$ dex ($\sigma=0.050$ dex, 7 stars), where the
first and second error bars refer to statistical and systematic errors,
respectively.

The abundance ratios for proton-capture elements are given in
Table~\ref{tablight} for UVES spectra, together with number of lines used and
rms scatter. All O abundances are detections; no measure of O abundance in star
5634-13 was possible because the forbidden line [O~I] 6300.31~\AA \ was affected
by sky contamination. For stars 5634-2 and 5634-9 only upper limits could be
measured for Al.

Abundances of $\alpha-$capture, iron-peak and neutron-capture elements for the
individual stars are listed in Table~\ref{tabalpha}, Table~\ref{tabiron}, 
and Table~\ref{tabneutron}, respectively, for the seven stars with UVES spectra.
Except for iron, all the abundances derived for stars observed with GIRAFFE are
listed in Table~\ref{tabgiraffe}. For these 35 stars all the element ratios are
referred to the average  iron abundance [Fe/H]$=-1.87$ dex.
Finally, in Table~\ref{meanabu} the mean abundances in NGC~5634 are summarized.

We  cross-matched our stars with the nine objects in Dias et al. (2016) and found
only three in common. Stars \# 9, 54, 87 in our sample are within 1\arcsec \ of
stars \# 11, 8, 7 in their list and have reasonably similar $V$ magnitudes
(theirs are instrumental values) and atmospheric parameters. We do not
dwell on the comparison since their results are based on lower resolution
spectra. However, their cluster average metallicity agrees with ours within the error bars and
their Mg abundance and [$\alpha$/Fe] are in very good accord.

We also have three stars in common with the APOGEE survey (Holtzman et al.
2015); however parameters were presented only for two of the stars in DR12.
Furthermore, a straightforward comparison is difficult, because of different
model atmospheres, spectral ranges and line list, and adopted methods.

Limiting the comparison to the optical range, 
ours is the second high-resolution optical spectroscopic study of this cluster.
S15 used a full spectroscopic
parameter determination for the two stars they analyzed in NGC~5634. Had we used
their atmospheric parameters for star 5634-2, in common with that study and observed
with similar resolution and wavelength coverage, we
would have obtained on average higher abundances by $0.003\pm 0.016$ dex
($\sigma=0.060$ dex) from 14 neutral species and lower by $-0.049\pm 0.022$ dex
($\sigma=0.058$ dex) from 7 singly ionised species, once our solar reference 
abundances from Gratton et al. (2003) are homogeneously adopted.

NGC~5634 seems to be an homogeneous cluster, as far as most elements are concerned.
By comparing the expected observational uncertainty due to errors affecting the
analysis (the internal errors in Table~\ref{sensu}) with the observed 
dispersion (the standard deviation about the mean values in Table~\ref{meanabu})
we can evaluate whether there is an intrinsic or cosmic scatter for a given
element among stars in NGC~5634. This exercise was made using the more accurate
abundances from UVES spectra; it is essentially equivalent to
compute the ``spread ratio" introduced by Cohen (2004) and it shows that 
there is evidence for an intrinsic spread only for O, Na, Al, and for a
possible spread for Fe, Y, and Eu.

However, we do not consider the spread in Fe to be real, since in this case the
observational uncertainty is very small due to the large number of measured
lines in UVES spectra. On the other extreme, when the abundance of a given specie is based on a
very limited number of lines, the spread ratio may be biased high (see Cohen
2004 for a discussion), as in the case of Y and Eu. Moreover, all five
determinations of Eu abundances from the line at 6437~\AA\ are upper limits.
Had we considered only the detections (from the Eu 6645~\AA\ line), the spread
ratio would be $< 1$, implying no intrinsic spread for Eu in NGC~5634.

In conclusion, the overall pattern of the chemical composition shows that 
NGC~5634 is a normal GC, where most elements do not present any
intrinsic star-to-star variation, except for those  involved in
proton-capture reactions in H burning at high temperature, such as O, Na, Al,
and as commonly observed 
in GCs (see Gratton et al. 2012).

\subsection{The light elements \label{lightabu}}

\begin{figure*}
\centering
\includegraphics[scale=0.52]{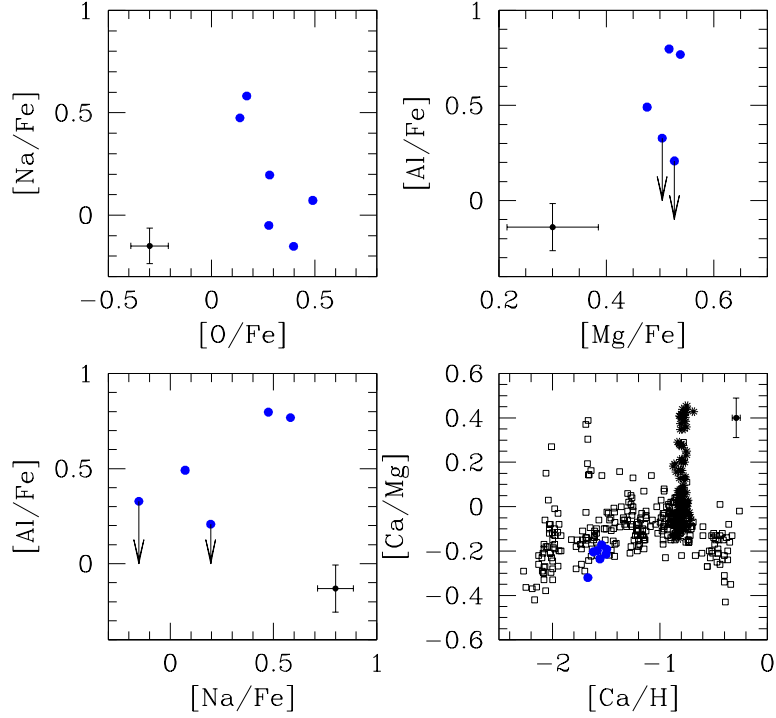}
\caption{Relations among proton-capture elements in stars of NGC~5634 
observed with UVES (filled
blue circles). Empty squares in the lower-right panel are stars in 24 GCs from
Carretta et al. (2009a,b), Carretta et al. (2010b,c), Carretta et al. (2011),
Carretta et al. (2013), Carretta et al. (2014b), Carretta et al. (2015), and
asterisks are stars in NGC~2808 from Carretta (2015). In each panel the error
bars represent internal errors.}
\label{anti}
\end{figure*}

The relations among proton-capture elements in stars of NGC~5634 
are summarized in Fig.~\ref{anti} for the UVES sample. 
Abundances of Na are anticorrelated with O abundances also in NGC~5634 (upper left panel), 
confirming the findings by S15 with a  sample three time larger. This cluster
shares the typical Na-O anticorrelation, the widespread chemical signature of
multiple stellar populations in MW GCs. Within the important limitations of
small number statistics, two stars have distinctly larger Na content and lower O
abundance than the other four stars. 

According to the homogeneous criteria used by our group to define stellar
generations in GCs (see Carretta et al. 2009a), the two
stars with the highest Na abundance would be second generation stars of the
intermediate I component, while the
other four giants belong to the first generation in NGC~5634 and reflect the
primordial P fraction, with the typical pure nucleosynthesis by type II SNe (the
average abundances for these  stars are [Na/Fe]=0.017 dex, $\sigma=0.151$ dex
and [O/Fe]=0.361 dex, $\sigma=0.102$ dex). The star with only an
Na abundance would be also classified  as a first generation star, due to its low ratio
[Na/Fe]=+0.052 dex.

We note that the two stars analyzed by S15 show (anticorrelated) spreads in O
and Na exceeding the estimated observational uncertainties, an evidence that
prompted the authors to claim an Na-O anticorrelation in NGC~5634.
Yet both stars would fall among our first generation stars, either if the
original abundances or those corrected for different solar abundances are
adopted. In our data, the observed spread in O for the stars of the P component
is comparable with the expected uncertainty due to the analysis, whereas the
spread in Na formally corresponds to 1.7$\sigma$. We cannot totally exclude a
certain amount of intrinsic spread among proton-capture elements in the first
generation  stars in NGC~5634, supporting previous findings by S15 based
on the Na abundances of their two stars, photometrically very similar.

The impression is supported by the upper-right and lower-left panels in
Fig.~\ref{anti}: no significant star-to-star variation is obser ved for Mg, while
the Al abundances cover a relatively large range which is only a lower limit
since for two P stars only upper limits to the Al abundances could be measured.

The lack of any extreme abundance variation in Mg is strongly confirmed by the
lower-right panel, where we compare abundances of Ca and Mg in NGC~5634 with
the values found in our homogeneous survey in 25 GCs. The stars analyzed in
NGC~5634 follow the trend of cluster stars with no extreme Mg depletion 
(see also S15).

To have a more robust estimate of the fraction of first and second generation stars
in NGC~5634 we may resort to the large statistics provided by Na abundances of
the combined UVES+GIRAFFE semple. We cannot distinguish second generation stars
with intermediate and extreme composition (this would require knowledge also of
O abundances), but the definition of the primordial P fraction in a GC simply
rests on Na abundances (Carretta et al. 2009a).  

\begin{figure*}
\centering
\includegraphics[bb=20 436 587 712,clip,scale=0.72]{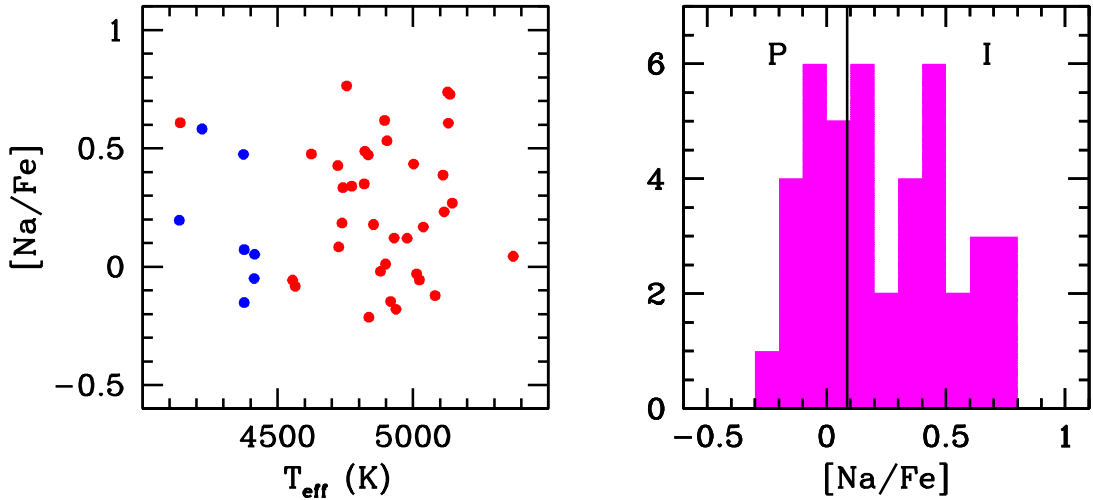}
\caption{Left-hand panel: [Na/Fe] ratios as a function of the temperature in
stars observed with UVES (blue squares) and GIRAFFE (red circles). Right-hand panel:
histogram of the [Na/Fe] ratios, where the line indicates the division between P
and I stars, based on our usual separation at [Na/Fe]$_{min}$+0.3.}
\label{m34na}
\end{figure*}

Using the total sample and separating at [Na/Fe]$_{min}$+0.3 dex, the fraction of P stars in NGC~5634 is $38 \pm 10\%$,
while the fraction of the second generation stars is $62 \pm 12\%$, where the
associated errors are due to Poisson statistics. These fraction are not very
different from the average of what found in most GCs (see e.g. Carretta et
al. 2009a, 2010a).

\subsection{Elements up to iron-peak \label{otherabu}}

\begin{figure}
\centering
\includegraphics[scale=0.42]{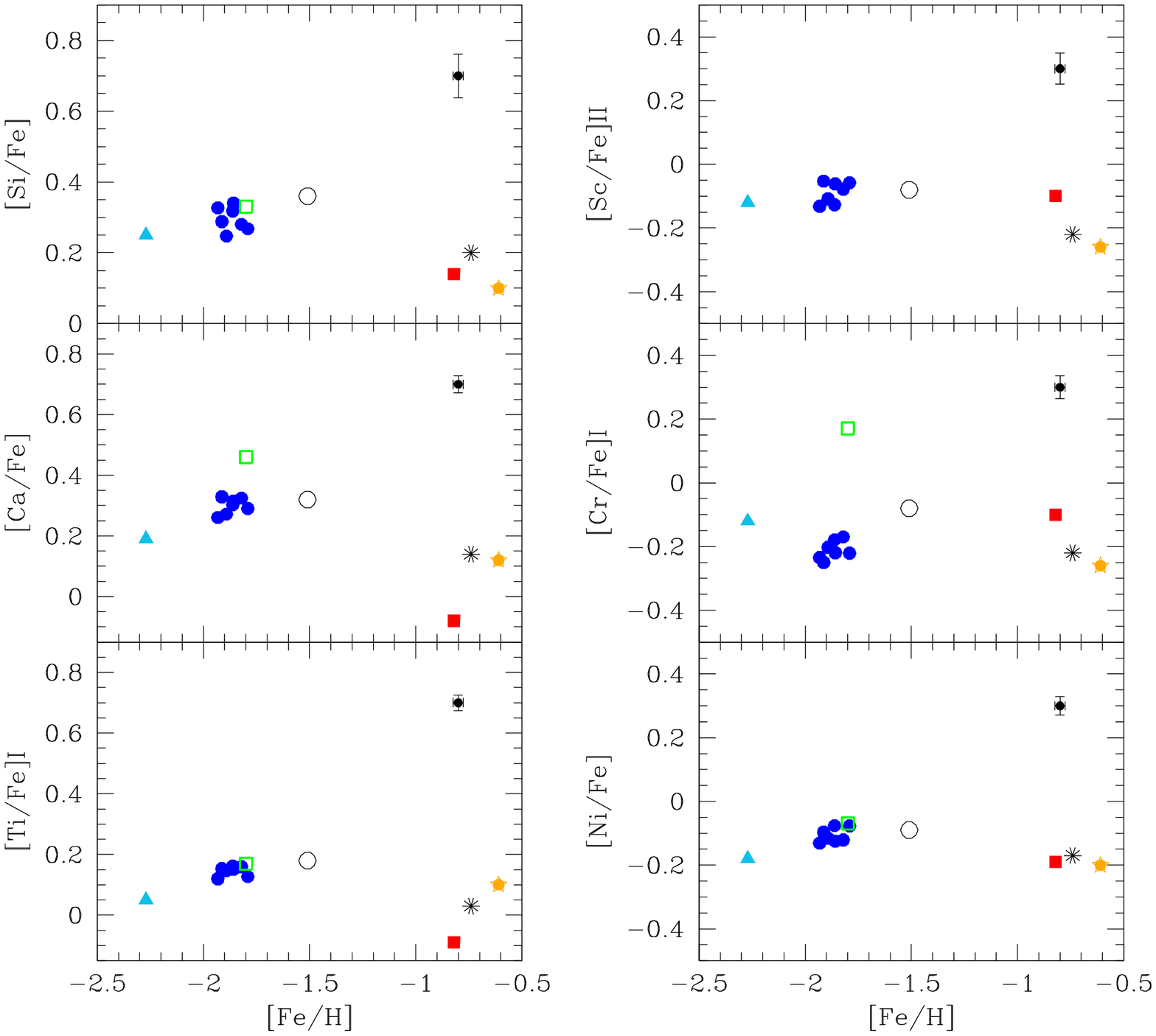}
\caption{Abundance ratios of $\alpha-$ and iron-peak elements derived from UVES
spectra in NGC~5634 as a function of the metallicity (filled blue circles). Also 
plotted are the average values from Table 11 of Carretta et al. (2014a) relative to the
Sgr nucleus (black asterisk, Carretta et al. 2010b), to M~54 (black empty
circle, Carretta et al. 2010b), Pal~12 (red filled square, Cohen 2004),
Terzan~7 (orange filled star, Sbordone et al. 2007), Terzan~8 (filled light-blue
triangle, Carretta et al. 2014a), and Arp~2 (green empty square, Mottini et al.
2008). Internal error bars refer to our UVES sample.}
\label{u34elefe}
\end{figure}

In Fig.~\ref{u34elefe} we summarize the chemical composition of NGC~5634 as
far as $\alpha-$elements and iron-peak elements are concerned,
using the individual values derived for stars with UVES spectra, as a function
of the metallicity.
As a comparison, we also plot the average value relative to the Sgr nucleus 
([Fe/H]=-0.74 dex,  Carretta et al. 2010b) and to the five GCs confirmed members of the Sgr dwarf 
galaxy: M~54 ([Fe/H]=-1.51 dex, Carretta et al. 2010b), Terzan~8 
([Fe/H]=-2.27 dex, Carretta et al. 2014a), Arp~2 ([Fe/H]=-1.80 dex, Mottini et 
al. 2008), Pal~12 ([Fe/H]=-0.82 dex, Cohen 2004), and Terzan~7 
([Fe/H]=-0.61 dex, Sbordone et al. 2007). In the last three cases, the values
were corrected to the scale of solar abundances presently used (Gratton et al.
2003).

Abundances of stars in NGC~5634 seem to be in good agreement with the 
metal-poor GCs associated to Sgr. Of course, by itself this cannot be a clearcut
proof of the membership of NGC~5634 to Sgr rather than to the Milky Way, since
at this low metal abundance the overall chemical pattern of the $\alpha-$ and
iron-peak elements simply reflects the typical floor of elemental abundances
established by the interplay of core-collapse and type Ia supernovae (see e.g.
Wheeler et al. 1989). This pattern is supported also by the abundances derived
from GIRAFFE spectra (Table~\ref{meanabu}). In Sect.~\ref{consgr} we will
discuss further elements to support the connection with Sgr dSph.

\subsection{Neutron-capture elements \label{neutron}}

Abundances for six neutron-capture elements in the UVES 
sample are listed in Table~\ref{tabneutron} and their cluster means are
in Table~\ref{meanabu}.
Ba~{\sc ii} and Nd~{\sc ii} lines could be treated as single unblended absorbers, so they
were treated with EW analyses.
Transitions of the other four neutron-capture elements had complications 
due to blending, hyperfine, or isotopic substructure and so were subjected
to synthetic spectrum analyses.
The elemental means appear to be straightforward:  the light n-capture element
Y and Zr are slightly under- and over-abundant, respectively, with respect to Fe, the traditional
s-process rare-earth elements Ba and La have solar abundances, the 
r-process dominant Eu is overabundant by about a factor of four, and the
r-/s- transition element Nd is overabundant by nearly a factor of three.
Here we comment on a few aspects of our derived n-capture abundances.

Our derived Ba and Eu abundances are in reasonable accord
with those reported by S15.
Barium is not the optimal element to assess the abundances of the low-Z end
of the rare earths, because all Ba~{\sc ii} lines are strong and thus less
sensitive to abundance than are weaker lines.  
Fortunately, abundances derived from weak La~{\sc ii} transitions are not in 
severe disagreement with the Ba values.

We determined Y abundances for our UVES sample from the Y~{\sc ii}  line at
5200.4~\AA, for homogeneity with what we did for Ter~8 and other GCs in our
sample (e.g., Carretta et al. 2014a). The cluster mean abundance from this
transition is [Y/Fe] = $-0.083$  ($\sigma = 0.119$, 7 stars; see
Table~\ref{tabneutron}).   The star-to-star scatter is relatively large, but
almost all stars have [Y/Fe]  $\lesssim$ 0.   Sbordone et al. (2015) reported
much lower Y abundance in NGC 5634:  [Y/Fe] = $-$0.40 from Star 2 and $-$0.33 from
Star 3.  

Star 2 in our study yielded [Y/Fe] = $-0.077$ from the synthesis 
of the 5200~\AA\ line.  
To investigate this $+$0.32 dex offset in [Y/Fe] with respect to 
S15, we synthesized other Y~{\sc ii} lines in Star~2 with the line 
analysis code MOOG (Sneden 1973).
Using the transition probabilities of Hannaford et al. (1982) and
Bi{\' e}mont et al. (2011), we used 10 Y~{\sc ii} lines to derive 
$<$[log~$\epsilon$(Y)]$>$ = $+$0.25 $\pm$ 0.03 ($\sigma$ = 0.08 dex).  
Adopting log~$\epsilon$(Y)$_\odot$ = 2.21 (Asplund et al. 2009) leads to a 
mean value of [Y/H] = $-$1.96 or [Y/Fe] = $-$0.09 (with the cluster mean 
[Fe/H] = $-$1.87), in good agreement with the value derived from the 
5200~\AA\ line alone.  

\begin{figure}
\centering
\includegraphics[bb= 42 146 566 590, clip, scale=0.42]{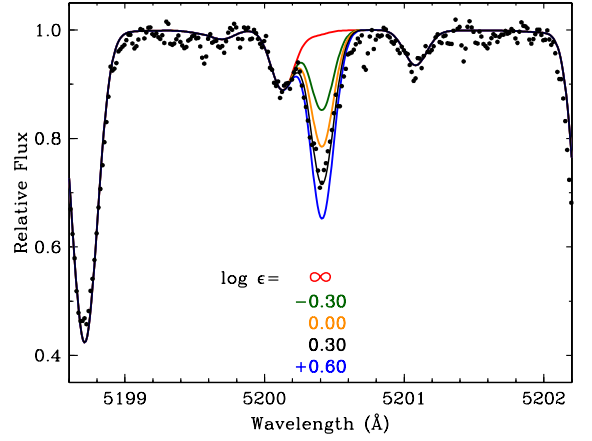} 
\caption{Synthetic spectra for the Y~{\sc ii} line 5200.4~\AA. Filled circles
indicate the observed spectrum of star 5634-2.} 
\label{f:spec5200}
\end{figure}

For Star~2 we display our synthetic/observed spectrum match in
Fig.~\ref{f:spec5200}. From this line our best estimate is log~$\epsilon$(Y) =
$+$0.30. We conclude that the spectrum synthesis of the line 5200~\AA\ alone
allows us to derive a fairly good estimate of the Y content. In general, we do not
find substantial Y deficiencies for any of our UVES sample in NGC~5634.

As suggested by the referee, we used the line profile for each 
synthesized line of Y~{\sc ii} published by S15 for star 2 and we fitted them
with our code and linelist. We obtained very consistent abundances from these
four lines ($\log \epsilon = +0.21 \pm 0.04$ dex, $\sigma=0.09$ dex and
$\log \epsilon = +0.28 \pm 0.03$ dex, $\sigma=0.06$ dex, with and without
continuum scattering). This exercise would give an offset 0.43-0.50 dex with
respect to the value of S15. At present, we cannot provide an explanation for
the difference.

\begin{figure}
\centering
\includegraphics[scale=0.42]{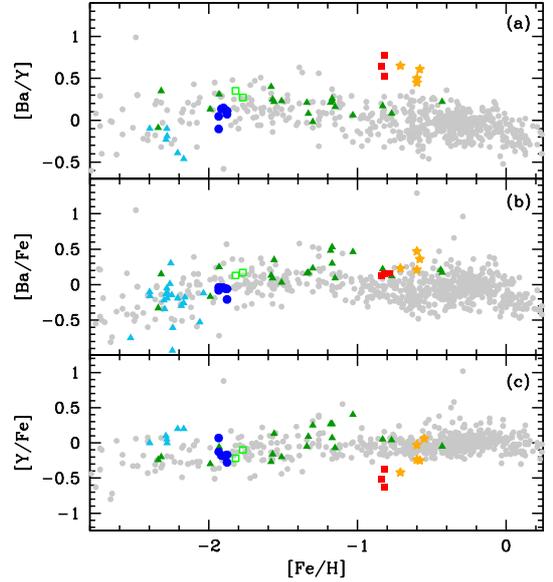} 
\caption{Abundance ratios of neutron-capture elements Y and Ba as a function of
metallicity. Grey filled circles are Galactic field stars from the compilation
by Venn et al. (2004). The other symbols are as in Fig.~\ref{u34elefe}:
Pal~12 (red filled squares, 
Cohen 2004), Terzan~7 (orange filled stars, Sbordone et al. 2007), Terzan~8
(filled light-blue triangles, Carretta et al. 2014a), and Arp~2 (green empty
squares, Mottini et al. 2008).}
\label{f:bay34}
\end{figure}

Derived Ba and Y abundances in NGC~5634 are in good agreement with the trend defined
by Galactic field stars of similar (low) metal abundances, as shown in
Fig.~\ref{f:bay34}. Again, the metal-rich GCs associated to the Sgr dwarf stand
out with respect to the Galactic field stars, while NGC~5634 cannot be
distinguished from its chemical composition alone, as also occurs for Terzan~8,
the classical metal-poor globular cluster of the family of GCs in Sgr. 

Here we
are comparing our NGC~5634 abundance ratios with those of other Sgr
clusters.  In the next section we will consider the kinematics of this
cluster in order to more firmly establish its membership in the Sgr
family.

\section{Connection with the Sagittarius dSph and internal kinematics \label{consgr}} 

\begin{figure}
\centering
\includegraphics[scale=0.42]{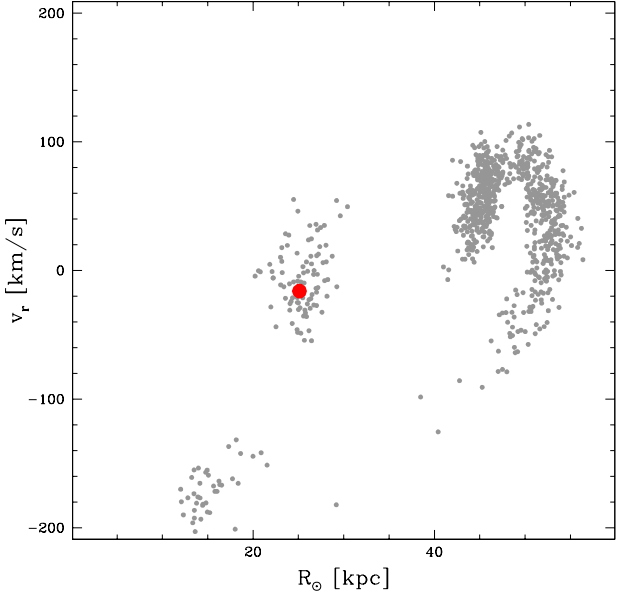}
\caption{Position and velocity of NGC~5634 (large filled red dot) compared to a
Sgr model (see text for details).}
\label{sgr}
\end{figure}

As introduced in Sect.~\ref{intro}, NGC~5634 is suggested to have formed in the
Sgr dSph and have then accreted by the Milky Way during the
subsequent tidal disruption of its host galaxy (Bellazzini et al. 2002; Law \&
Majewski 2010a). In this scenario, this GC is expected to follow an orbit which
is similar to that of its parent galaxy and to be embedded in its stream. We
checked this hypothesis by comparing the position and systemic motion of  NGC~5634 
with the prediction of the Sagittarius model 
by Law \& Majewski (2010b). Among the three models
presented in that paper we used the one assuming a prolate (q=1.25) Galactic
halo which provides a better fit to the observational constraints on the
location and kinematics of the Sagittarius stream stars. The distribution in the
heliocentric distance-radial velocity plane of 
particles within $5\deg$ from the present-day position of NGC~5634 is shown in
Fig. \ref{sgr}. It is apparent that in this diagram NGC~5634 lies well within a clump of
Sagittarius particles. These particles belong to the trailing arm of its stream
which have been lost by the satellite between 3 and 7 Gyr ago. This is
consistent with what found by B02 and Law \& Majewski
(2010a) even if they used a slightly different systemic radial velocity for this
cluster  (-44.4 km~s$^{-1}$).

\begin{figure}
\centering
\includegraphics[scale=0.42]{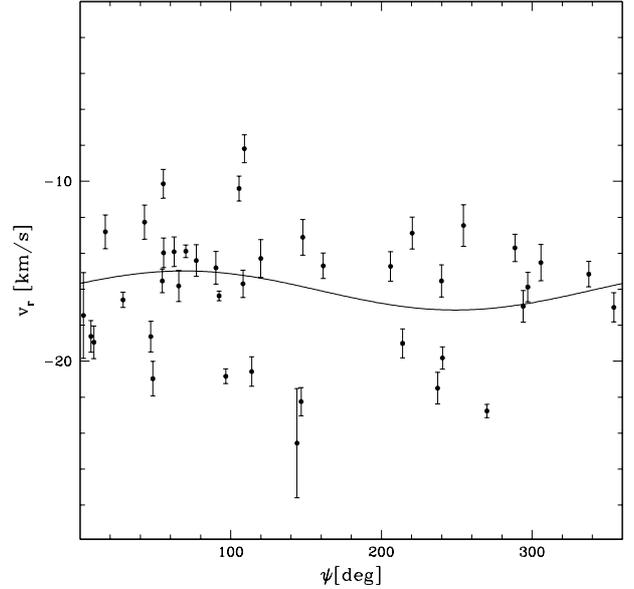}
\caption{Systemic rotation of NGC~5634: the best-fit solution is compatible with
negligible  rotation.}
\label{rot}
\end{figure}

\subsection{Internal kinematics}

The present data represents the most extensive set of radial velocities for
NGC 5634 and can be useful to study the internal kinematics of this cluster.
As a first step, we test the presence of systemic rotation. For this purpose, 
in Fig. \ref{rot} the RVs of the 42 {\it bona fide} members are
plotted against their position angles. The best-fit sinusoidal curve indicates a rotation amplitude of
$A_{rot}~sin i=1.08\pm1.34$ km~s$^{-1}$, compatible with no significant
rotation.

We then used our RV dataset to estimate the dynamical mass of the
system. For this purpose we fitted the distribution of RVs with a
set of single and multimass King-Michie models (King 1966; Gunn \& Griffin
1979). In particular, for each model we tuned the model mass to maximize the
log-likelihood 
$$L=-\sum_{i=1}^{N}\left(\frac{(v_{i}-\langle v
\rangle)^{2}}{(\sigma_{i}^{2}+\epsilon_{i}^{2})}+ln
(\sigma_{i}^{2}+\epsilon_{i}^{2})\right)$$
where N(=42) is the number of available RVs, $v_{i}$ and $\epsilon_{i}$ are 
the radial velocity of the $i$-th star and its associated uncertainty and
$\sigma_{i}$ is the line-of-sight velocity dispersion predicted by the model at
the distance from the cluster centre of the $i$-th star. 
The best-fit single mass model provides a total mass of $1.64\times10^{5} M_{\odot}$.
This quantity has to be considered however a lower limit to the actual total 
mass because of the effect of two-body relaxation affecting this estimate. 
Indeed, RGB stars
(i.e. those for whom RVs are available) are on average more
massive than typical cluster stars and are therefore expected to become kinematically 
colder and more concentrated than the latter after a timescale comparable to the
half-mass relaxation time. To account for this effect we performed the same
analysis adopting a set of multimass models with various choices of the
present-day mass function. In particular, we adopted a power-law mass function
with indices ranging from -2.35 (i.e. Salpeter 1955) to 0. We adopted the
prescriptions for dark remnants and binaries of Sollima et al. (2012) assuming a
binary fraction of 10\% and a flat distribution of mass-ratios.
The derived masses turn out to be 1.72, 1.82, 2.04 and 1.93$\times10^{5} M_{\odot}$ for
mass function slopes of $\alpha=$0, -1, -2 and -2.35, respectively, with a
typical uncertainty of $\sim 4.5\times10^{4} M_{\odot}$.
By assuming the
absolute magnitude $M_{V}=-7.69$ listed for this cluster in the Harris catalog 
(1996, 2010 edition) the corresponding M/L ratio are 1.68, 1.79, 2.01 and 1.90. 
Significantly larger M/L ratios (by a factor 1.24) are 
instead obtained if the integrated magnitude by McLaughlin \& van der Marel
(2005) is adopted.

\section{Summary and conclusions \label{summary}}

As part of our homogeneous study of GCs, we obtained FLAMES spectra of 42 member
stars in the cluster NGC~5634,  associated to the Sgr dSph, in particular to one
of  the arms if  the Sgr stream. We measured elemental abundances for many
elements in the UVES spectra of seven RGB stars and of several elements in the
GIRAFFE spectra of 35 (mainly) RGB and AGB stars (see Fig.\ref{cmdoss}). 

We found clear evidence of multiple stellar populations in this cluster, 
indicated by the classical
(anti)correlations between light elements (O, Na; Mg, Al). Although we could not
characterize completely these anticorrelations because we lack oxygen abundance
for all GIRAFFE targets, we can divide the observed stars into primordial P and
Intermediate I fractions (Carretta et al. 2009a,b) simply using their Na abundance.
Apparently, for this Sgr cluster the fraction of first and second generation
stars is not dramatically different from typical values found in MW GCs from
spectroscopy (e.g. Carretta et al. 2009a,2010a, Johnson \& Pilachowski 2012).

We support the connection between NGC~5634 and the Sgr dSph (B02, Law \&
Majewski 2010a) both on the basis of the cluster RV and position and on the
chemical abundances. The second evidence is not however clearcut, since NGC~5634
resembles both MW and low-metallicity Sgr GCs in $\alpha$ and
neutron-capture elements.
We do not confirm the very low Y {\sc ii} abundance found by S15.

This study adds yet another confirmation of the ubiquitous presence of
light-element anticorrelations, i.e., multiple populations, among old and
massive GCs, independent of their formation place. Metal-poor GCs apparently
formed following the same chain of events regardless their birth environment was
the central Galaxy (Milky Way) or its most prominent disrupting dwarf satellite
(Sgr).

\begin{table*}
\centering
\caption{Sensitivities of abundance ratios to variations in the atmospheric
parameters and to errors in the equivalent widths, and errors in abundances for
stars of NGC~5634 observed with UVES.}
\begin{tabular}{lrrrrrrrr}
\hline
Element     & Average  & T$_{\rm eff}$ & $\log g$ & [A/H]   & $v_t$    & EWs     & Total   & Total      \\
            & n. lines &      (K)      &  (dex)   & (dex)   &kms$^{-1}$& (dex)   &Internal & Systematic \\
\hline        
Variation&             &  50           &   0.20   &  0.10   &  0.10    &         &         &            \\
Internal &             &   5           &   0.04   &  0.05   &  0.07    & 0.01    &         &            \\
Systematic&            &  42           &   0.06   &  0.07   &  0.03    &         &         &            \\
\hline
$[$Fe/H$]${\sc  i}& 70 &    +0.074     & $-$0.013 &$-$0.015 & $-$0.023 & 0.014  &0.024    &0.065	\\
$[$Fe/H$]${\sc ii}& 13 &  $-$0.029     &   +0.003 &  +0.001 &   +0.006 & 0.033  &0.034    &0.026	\\
$[$O/Fe$]${\sc  i}&  2 &  $-$0.055     &   +0.088 &  +0.042 &	+0.020 & 0.085  &0.091    &0.077	\\
$[$Na/Fe$]${\sc i}&  2 &  $-$0.035     & $-$0.039 &$-$0.015 &	+0.019 & 0.085  &0.087    &0.107	\\
$[$Mg/Fe$]${\sc i}&  2 &  $-$0.021     & $-$0.018 &$-$0.004 & $-$0.002 & 0.085  &0.085    &0.023	\\
$[$Al/Fe$]${\sc i}&  1 &  $-$0.006     &   +0.015 &  +0.021 &	+0.040 & 0.120  &0.124    &0.132	\\
$[$Si/Fe$]${\sc i}&  4 &  $-$0.060     &   +0.027 &  +0.015 &	+0.019 & 0.060  &0.062    &0.053	\\
$[$Ca/Fe$]${\sc i}& 19 &  $-$0.008     & $-$0.013 &$-$0.007 & $-$0.002 & 0.028  &0.028    &0.013	\\
$[$Sc/Fe$]${\sc ii}& 6 &    +0.028     & $-$0.004 &  +0.004 &   +0.003 & 0.049  &0.049    &0.027	\\
$[$Ti/Fe$]${\sc i}& 23 &    +0.037     & $-$0.012 &$-$0.013 &	+0.000 & 0.025  &0.026    &0.032	\\
$[$Ti/Fe$]${\sc ii}& 9 &    +0.022     & $-$0.015 &$-$0.003 & $-$0.008 & 0.040  &0.041    &0.021	\\
$[$V/Fe$]${\sc i} &  9 &    +0.030     & $-$0.016 &$-$0.008 &	+0.017 & 0.040  &0.042    &0.028	\\
$[$Cr/Fe$]${\sc i}& 12 &    +0.022     & $-$0.015 &$-$0.013 & $-$0.007 & 0.035  &0.036    &0.022	\\
$[$Cr/Fe$]${\sc ii}& 5 &    +0.000     & $-$0.014 &$-$0.011 &	+0.006 & 0.054  &0.054    &0.020	\\
$[$Mn/Fe$]${\sc i}&  6 &    +0.016     & $-$0.010 &$-$0.007 &	+0.005 & 0.049  &0.049    &0.015	\\
$[$Co/Fe$]${\sc i}&  3 &  $-$0.016     & $-$0.001 &  +0.003 &	+0.019 & 0.069  &0.071    &0.017	\\
$[$Ni/Fe$]${\sc i}& 18 &  $-$0.007     &   +0.011 &  +0.006 &	+0.009 & 0.028  &0.029    &0.011	\\
$[$Cu/Fe$]${\sc i}&  1 &    +0.012     &   +0.005 &  +0.001 &   +0.003 & 0.120  &0.120    &0.041	\\
$[$Zn/Fe$]${\sc i}&  1 &  $-$0.093     &   +0.048 &  +0.022 &   +0.005 & 0.120  &0.121    &0.081	\\
$[$Y/Fe$]${\sc ii}&  1 &    +0.035     & $-$0.014 &$-$0.001 & $-$0.007 & 0.120  &0.120    &0.054	\\
$[$Zr/Fe$]${\sc ii}& 1 &    +0.129     &   +0.090 &  +0.100 &	+0.109 & 0.120  &0.152    &0.128	\\
$[$Ba/Fe$]${\sc ii}& 3 &    +0.047     & $-$0.006 &  +0.002 & $-$0.071 & 0.069  &0.085    &0.089	\\
$[$La/Fe$]${\sc ii}& 1 &    +0.046     & $-$0.006 &  +0.005 &	+0.009 & 0.085  &0.085    &0.046	\\
$[$Nd/Fe$]${\sc ii}& 4 &    +0.046     & $-$0.010 &  +0.003 &   +0.003 & 0.060  &0.060    &0.045	\\
$[$Eu/Fe$]${\sc ii}& 2 &    +0.031     & $-$0.001 &  +0.007 &	+0.007 & 0.085  &0.085    &0.054	\\
\hline
\end{tabular}
\label{sensu}
\end{table*}

\begin{table*}
\centering
\caption[]{Sensitivities of abundance ratios to variations in the atmospheric
parameters and to errors in the equivalent widths, and errors in abundances for
stars of NGC~5634 observed with GIRAFFE.}
\begin{tabular}{lrrrrrrrr}
\hline
Element     & Average  & T$_{\rm eff}$ & $\log g$ & [A/H]   & $v_t$    & EWs     & Total   & Total      \\
            & n. lines &      (K)      &  (dex)   & (dex)   &kms$^{-1}$& (dex)   &Internal & Systematic \\
\hline        
Variation&             &  50           &   0.20   &  0.10   &  0.10    &         &         &            \\
Internal &             &   5           &   0.04   &  0.09   &  0.83    & 0.06    &         &            \\
Systematic&            &  61           &   0.06   &  0.06   &  0.14    &         &         &            \\
\hline
$[$Fe/H$]${\sc  i}& 6 &    +0.044  & $-$0.008 &$-$0.008 & $-$0.010 & 0.055  &0.100 &0.058 \\
$[$Na/Fe$]${\sc i}& 2 &  $-$0.022  & $-$0.018 &  +0.002 &   +0.007 & 0.095  &0.112 &0.056 \\
$[$Mg/Fe$]${\sc i}& 1 &  $-$0.016  &   +0.002 &  +0.002 &   +0.005 & 0.135  &0.141 &0.023 \\
$[$Si/Fe$]${\sc i}& 3 &  $-$0.026  &   +0.017 &  +0.007 &   +0.008 & 0.078  &0.103 &0.035 \\
$[$Ca/Fe$]${\sc i}& 1 &  $-$0.010  & $-$0.002 &  +0.001 & $-$0.003 & 0.135  &0.137 &0.017 \\
$[$Sc/Fe$]${\sc ii}&5 &  $-$0.036  &   +0.081 &  +0.023 &   +0.002 & 0.060  &0.068 &0.051 \\
$[$V/Fe$]${\sc i} & 2 &    +0.027  & $-$0.007 &$-$0.004 &   +0.007 & 0.095  &0.112 &0.038 \\
$[$Ni/Fe$]${\sc i}& 2 &    +0.015  &   +0.005 &  +0.003 &   +0.001 & 0.095  &0.096 &0.021 \\
\hline
\end{tabular}
\label{sensm}
\end{table*}

\begin{table*}
\centering
\caption[]{Light element abundances.}
\begin{tabular}{rccccrcccrcrc}
\hline
Star   &   nr &  [O/Fe]{\sc i} &  rms &  nr &  [Na/Fe]{\sc i} &  rms &  nr &  [Mg/Fe]{\sc i} &  rms &  nr &  [Al/Fe]{\sc i} &  rms  \\
\hline
 2 &2 & 0.281 &0.029 &3 & 0.196 &0.057 &3 &0.527 &0.052 &1 &$<$0.208 &      \\
 5 &2 & 0.170 &0.080 &3 & 0.582 &0.060 &2 &0.538 &0.066 &1 &0.768 &      \\
 7 &2 & 0.490 &0.035 &2 & 0.072 &0.013 &2 &0.476 &0.182 &2 &0.491 &0.025 \\
 9 &1 & 0.397 &      &2 &-0.152 &0.061 &3 &0.504 &0.061 &1 &$<$0.328 &      \\
10 &1 & 0.137 &      &2 & 0.475 &0.046 &3 &0.517 &0.120 &2 &0.797 &0.002 \\
12 &1 & 0.277 &      &1 &-0.050 &      &2 &0.581 &0.022 &  &	  &      \\
13 &  &       &      &1 & 0.052 &      &2 &0.487 &0.071 &  &	  &      \\
\hline
\end{tabular}
\label{tablight}
\end{table*}

\begin{table*}
\centering
\caption[]{$\alpha$ element abundances.}
\begin{tabular}{rccccrcccrccc}
\hline
Star   &   nr &  [Si/Fe]{\sc i} &  rms &  nr &  [Ca/Fe]{\sc i} &  rms &  nr &  [Ti/Fe]{\sc i} &  rms &  nr &  [Ti/Fe]{\sc ii} &  rms  \\
\hline
 2 &4 &0.288 &0.138 &18 &0.329 &0.143 &25 &0.153 &0.166 & 9 &0.147 &0.090\\
 5 &4 &0.318 &0.074 &19 &0.303 &0.115 &25 &0.162 &0.140 & 9 &0.190 &0.083\\
 7 &3 &0.247 &0.130 &19 &0.272 &0.085 &25 &0.147 &0.106 &11 &0.197 &0.151\\
 9 &5 &0.268 &0.047 &20 &0.290 &0.085 &24 &0.127 &0.111 &11 &0.139 &0.159\\
10 &4 &0.280 &0.059 &19 &0.325 &0.084 &25 &0.160 &0.094 & 9 &0.183 &0.144\\
12 &3 &0.327 &0.105 &19 &0.261 &0.116 &19 &0.120 &0.140 & 8 &0.150 &0.176\\
13 &2 &0.340 &0.135 &20 &0.315 &0.167 &18 &0.152 &0.137 & 9 &0.171 &0.164\\
\hline
\end{tabular}
\label{tabalpha}
\end{table*}

\begin{table*}
\centering
\setlength{\tabcolsep}{1mm}
\caption[]{Iron-peak abundances.}
\tiny
\begin{tabular}{rccccrcccrcccrcccrcccrcccrccc}
\hline
Star   
&   nr &  [Sc/Fe]{\sc ii} &  rms &  nr &  [V/Fe]{\sc i} &  rms &  nr &  [Cr/Fe]{\sc i} &  rms &  nr &  [Cr/Fe]{\sc ii} &  rms  &   nr &  [Mn/Fe]{\sc  i} &  rms &  nr &  [Co/Fe]{\sc i}&  rms &  nr &  [Ni/Fe]{\sc i} &  rms &  nr &  [Cu/Fe]{\sc i}  &  rms  &   nr &  [Zn/Fe]{\sc i} &  rms \\
\hline
 2 &5 &-0.053 &0.137 &11 &-0.131 &0.113 &13 &-0.250 &0.098 &6 &-0.009 &0.137 &6 &-0.486 &0.189 &3 &-0.038 &0.058 &21 &-0.097 &0.119 &1 &-0.597 &      &1 & 0.016  \\
 5 &6 &-0.127 &0.116 &12 &-0.100 &0.096 &14 &-0.179 &0.123 &5 & 0.046 &0.174 &7 &-0.495 &0.210 &3 & 0.000 &0.128 &22 &-0.076 &0.119 &1 &-0.498 &      &1 & 0.015  \\
 7 &7 &-0.109 &0.115 & 8 &-0.158 &0.085 &10 &-0.203 &0.061 &6 & 0.082 &0.122 &7 &-0.523 &0.163 &3 &-0.014 &0.218 &18 &-0.116 &0.100 &1 &-0.498 &      &1 &-0.063  \\
 9 &8 &-0.058 &0.132 &11 &-0.159 &0.080 &13 &-0.221 &0.082 &6 &-0.022 &0.094 &7 &-0.478 &0.217 &3 & 0.012 &0.063 &21 &-0.077 &0.100 &1 &-0.599 &      &1 &-0.002  \\
10 &8 &-0.078 &0.101 &11 &-0.146 &0.136 &12 &-0.170 &0.071 &3 &-0.080 &0.109 &7 &-0.520 &0.181 &3 &-0.027 &0.071 &20 &-0.121 &0.102 &1 &-0.609 &      &1 & 0.030  \\
12 &6 &-0.132 &0.069 & 7 &-0.098 &0.121 &10 &-0.235 &0.131 &3 &-0.001 &0.041 &4 &-0.488 &0.195 &2 & 0.030 &0.095 &11 &-0.131 &0.094 &1 &-0.798 &      &1 & 0.099  \\
13 &5 &-0.062 &0.145 & 3 &-0.113 &0.091 & 9 &-0.219 &0.102 &3 & 0.000 &0.060 &4 &-0.500 &0.138 &1 &-0.013 &	 &13 &-0.125 &0.113 &1 &-0.502 &      &1 &-0.005  \\
\hline
\end{tabular}
\label{tabiron}
\end{table*}

\begin{table*}
\centering
\caption[]{Neutron-capture abundances.}
\tiny
\begin{tabular}{rccccrcccccccrcccrcc}
\hline
Star   
&   nr &[Y/Fe]{\sc ii} &  rms &  nr &  [Zr/Fe]{\sc ii} &  rms &  nr &  [Ba/Fe]{\sc ii} &  rms &  nr &  [La/Fe]{\sc ii} &  rms  &  nr &  [Nd/Fe]{\sc ii}&  rms &  nr &  [Eu/Fe]{\sc ii} &  rms  \\
\hline
 2 &1 &$-$0.077 & & 1&$<$0.153 & &3 &-0.081 &0.026 &2 &+0.223 &0.030 &4 &+0.384 &0.084 &2 &+0.633 &0.071	\\
 5 &1 &$-$0.123 & & 1&   +0.227 & &3 &-0.056 &0.081 &2 &+0.027 &0.030 &4 &+0.456 &0.058 &1 &+0.487 &  	\\
 7 &1 &$-$0.125 & & 1&   +0.125 & &3 &-0.043 &0.057 &2 &+0.125 &0.030 &4 &+0.518 &0.078 &2 &+0.660 &0.035	\\
 9 &1 &$-$0.163 & & 1&   +0.087 & &3 &-0.208 &0.055 &2 &+0.087 &0.030 &4 &+0.380 &0.069 &1 &+0.447 &  	\\
10 &1 &$-$0.084 & & 1&   +0.046 & &3 &-0.059 &0.099 &2 &+0.146 &0.030 &4 &+0.432 &0.047 &2 &+0.456 &0.071	\\
12 &1 &$-$0.179 & & 1&   +0.071 & &3 &-0.041 &0.082 &2 &+0.071 &0.030 &4 &+0.450 &0.076 &2 &+0.631 &0.141	\\
13 &1 &   +0.173 & & 1&   +0.473 & &3 &-0.037 &0.111 &2 &+0.073 &0.030 &4 &+0.532 &0.055 &2 &+0.783 &0.212 	\\
\hline
\end{tabular}
\label{tabneutron}
\end{table*}

\begin{table*}
\centering
\setlength{\tabcolsep}{1mm}
\caption[]{Abundances from GIRAFFE spectra.}
\scriptsize
\begin{tabular}{rcrcccrcccrcccrcccrcccr}
\hline
Star   
&     nr &  [Na/Fe]{\sc i} &  rms &  nr &  [Mg/Fe]{\sc i} &  rms &  nr &  [Si/Fe]{\sc ii} &  rms  &   nr &  [Ca/Fe]{\sc  i} &  rms &  nr &  [Sc/Fe]{\sc ii}&  rms &  nr &  [V/Fe]{\sc i} &  rms &  nr & [Ni/Fe]{\sc i}  &  rms  \\
\hline
100& 2 &  0.178& 0.055& 1 & 0.464 &      &2 &  0.377& 0.022&1 &  0.345&      &5 &  0.064& 0.064&   &       &      &2 & -0.203& 0.120 \\ 
103& 2 &  0.532& 0.091& 1 & 0.445 &      &1 &  0.290&      &1 &  0.539&      &5 &  0.057& 0.096&   &       &      &2 & -0.189& 0.002 \\ 
113& 2 &  0.011& 0.036& 1 & 0.523 &      &2 &  0.438& 0.069&1 &  0.385&      &5 & -0.046& 0.081& 1 & -0.005&      &2 & -0.071& 0.139 \\ 
125& 1 &  0.121&      & 1 & 0.571 &      &4 &  0.446& 0.149&1 &  0.340&      &6 & -0.102& 0.154&   &       &      &2 & -0.095& 0.037 \\ 
132& 2 & -0.056& 0.085& 1 & 0.643 &      &4 &  0.407& 0.134&1 &  0.465&      &5 &  0.044& 0.090&   &       &      &2 & -0.084& 0.133 \\ 
135& 2 & -0.030& 0.061& 1 & 0.498 &      &1 &  0.284&      &1 &  0.389&      &3 & -0.011& 0.274&   &       &      &2 & -0.123& 0.258 \\ 
142& 1 &  0.434&      & 1 & 0.477 &      &2 &  0.413& 0.006&1 &  0.330&      &5 & -0.087& 0.102&   &       &      &2 & -0.120& 0.153 \\ 
146& 1 &  0.168&      & 1 & 0.507 &      &4 &  0.456& 0.149&1 &  0.357&      &4 & -0.095& 0.258&   &       &      &1 & -0.136&       \\ 
152& 1 &  0.120&      & 1 & 0.623 &      &2 &  0.414& 0.042&1 &  0.408&      &4 & -0.078& 0.048&   &       &      &2 & -0.142& 0.069 \\ 
163& 1 & -0.122&      & 1 & 0.520 &      &2 &  0.383& 0.022&1 &  0.320&      &3 & -0.024& 0.189&   &       &      &2 & -0.079& 0.027 \\ 
167& 1 &  0.387&      & 1 & 0.579 &      &  &       &      &1 &  0.301&      &5 &  0.000& 0.123&   &       &      &  &       &       \\ 
169& 1 &  0.607&      & 1 & 0.483 &      &3 &  0.410& 0.091&1 &  0.314&      &5 & -0.022& 0.133&   &       &      &2 & -0.039& 0.157 \\ 
173& 2 &  0.232& 0.055& 1 & 0.562 &      &1 &  0.430&      &1 &  0.405&      &4 & -0.038& 0.181&   &       &      &2 & -0.151& 0.028 \\ 
182& 1 &  0.738&      & 1 & 0.474 &      &  &       &      &1 &  0.405&      &3 &  0.006& 0.193&   &       &      &  &       &       \\ 
189& 2 &  0.728& 0.046& 1 & 0.561 &      &  &       &      &1 &  0.399&      &4 &  0.035& 0.051&   &       &      &  &       &       \\ 
194& 2 &  0.269& 0.024& 1 & 0.512 &      &  &       &      &1 &  0.524&      &4 & -0.010& 0.161&   &       &      &  &       &       \\ 
19 & 2 & -0.057& 0.030& 1 & 0.509 &      &6 &  0.402& 0.137&1 &  0.309&      &5 & -0.026& 0.040& 4 & -0.148& 0.151&3 & -0.120& 0.003 \\ 
22 & 2 & -0.083& 0.086& 1 & 0.421 &      &3 &  0.359& 0.093&1 &  0.268&      &5 & -0.014& 0.081& 1 & -0.121&      &2 &  0.018& 0.011 \\ 
30 & 2 &  0.350& 0.021& 1 & 0.490 &      &3 &  0.445& 0.248&1 &  0.337&      &5 &  0.026& 0.082& 2 & -0.121& 0.099&3 & -0.170& 0.208 \\ 
35 & 2 &  0.476& 0.017& 1 & 0.465 &      &5 &  0.435& 0.129&1 &  0.269&      &5 &  0.011& 0.075& 2 & -0.140& 0.044&2 & -0.127& 0.393 \\ 
3  & 2 &  0.608& 0.043& 1 & 0.523 &      &7 &  0.407& 0.070&1 &  0.335&      &5 &  0.022& 0.038& 4 & -0.061& 0.079&2 & -0.152& 0.024 \\ 
45 & 2 &  0.083& 0.058& 1 & 0.410 &      &5 &  0.467& 0.152&1 &  0.279&      &5 & -0.035& 0.078&   &       &      &2 & -0.058& 0.015 \\ 
46 & 2 &  0.427& 0.080& 1 & 0.323 &      &4 &  0.446& 0.254&1 &  0.285&      &6 & -0.070& 0.115&   &       &      &2 & -0.167& 0.049 \\ 
54 & 2 &  0.334& 0.062& 1 & 0.498 &      &4 &  0.374& 0.119&1 &  0.308&      &5 &  0.038& 0.071& 3 & -0.150& 0.141&2 & -0.134& 0.155 \\ 
58 & 2 &  0.764& 0.056& 1 & 0.559 &      &5 &  0.350& 0.112&1 &  0.409&      &5 &  0.030& 0.093& 1 & -0.165&      &3 & -0.172& 0.132 \\ 
60 & 2 &  0.184& 0.081& 1 & 0.540 &      &2 &  0.449& 0.126&1 &  0.334&      &6 & -0.061& 0.247&   &       &      &2 & -0.149& 0.168 \\ 
62 & 2 &  0.340& 0.081& 1 & 0.536 &      &4 &  0.379& 0.172&1 &  0.420&      &5 &  0.058& 0.080&   &       &      &2 & -0.049& 0.131 \\ 
64 & 2 & -0.180& 0.030& 1 & 0.432 &      &4 &  0.438& 0.185&1 &  0.373&      &5 &  0.058& 0.093&   &       &      &2 & -0.143& 0.134 \\ 
70 & 2 &  0.472& 0.033& 1 & 0.539 &      &5 &  0.419& 0.198&1 &  0.424&      &5 &  0.006& 0.067&   &       &      &2 & -0.138& 0.177 \\ 
72 & 2 & -0.020& 0.167& 1 & 0.478 &      &3 &  0.391& 0.206&1 &  0.397&      &6 & -0.102& 0.096&   &       &      &2 & -0.157& 0.202 \\ 
78 & 2 & -0.214& 0.019& 1 & 0.474 &      &4 &  0.389& 0.091&1 &  0.355&      &4 & -0.113& 0.219& 1 & -0.104&      &2 & -0.157& 0.077 \\ 
82 & 1 & -0.147&      & 1 & 0.388 &      &4 &  0.408& 0.224&1 &  0.267&      &  &       &      &   &       &      &1 & -0.185&       \\ 
87 & 2 &  0.488& 0.041& 1 & 0.424 &      &4 &  0.385& 0.150&1 &  0.334&      &5 &  0.009& 0.184&   &       &      &2 & -0.139& 0.273 \\ 
93 & 2 &  0.618& 0.040& 1 & 0.500 &      &4 &  0.449& 0.239&1 &  0.415&      &5 &  0.055& 0.096&   &       &      &2 &  0.080& 0.026 \\ 
97 & 1 &  0.044&      & 1 & 0.489 &      &1 &  0.413&      &1 &  0.440&      &5 &  0.045& 0.089&   &       &      &1 & -0.058&       \\ 
\hline
\end{tabular}
\label{tabgiraffe}
\end{table*}

\begin{acknowledgements}
We thank P. Montegriffo for his software CataPack, Michele Bellazzini
for useful comments, and the referee Luca Sbordone for his thoughful review.
Support for this work has come in
part from the US NSF grant AST-1211585.
This research has made use of Vizier and SIMBAD, operated at
CDS, Strasbourg, France, and NASA's Astrophysical Data System.
The Guide Star Catalogue-II is a joint project of the Space Telescope
Science Institute and the Osservatorio Astronomico di Torino. Space
Telescope Science Institute is operated by the Association of
Universities for Research in Astronomy, for the National Aeronautics
and Space Administration under contract NAS5-26555. The participation
of the Osservatorio Astronomico di Torino is supported by the Italian
Council for Research in Astronomy. Additional support is provided by
European Southern Observatory, Space Telescope European Coordinating
Facility, the International GEMINI project and the European Space
Agency Astrophysics Division.This publication makes use of data products
from the Two Micron All Sky Survey, which is a joint project of the University
of Massachusetts and the Infrared Processing and Analysis Center/California
Institute of Technology, funded by the National Aeronautics and Space
Administration and the National Science Foundation. 
\end{acknowledgements}

\begin{table}
\centering
\caption{Mean abundances in NGC~5634.}
\setlength{\tabcolsep}{1.5mm}
\begin{tabular}{lllllrcc}
\hline
Element              &  stars & mean &rms & Note & stars & mean &rms\\
\hline
    & \multicolumn{3}{c}{UVES} & & \multicolumn{3}{c}{GIRAFFE} \\
\hline
$[$O/Fe$]${\sc i}    & 6&    +0.292 & 0.134 & EW   &  &        &        \\
$[$Na/Fe$]${\sc i}   & 7&    +0.168 & 0.271 & EW   &35&  +0.252& 0.287  \\
$[$Mg/Fe$]${\sc i}   & 7&    +0.519 & 0.035 & EW   &35&  +0.498& 0.064  \\
$[$Al/Fe$]${\sc i}   & 5&    +0.498 & 0.293 & EW   &  &        &        \\
$[$Si/Fe$]${\sc i}   & 7&    +0.295 & 0.034 & EW   &31&  +0.405& 0.043  \\
$[$Ca/Fe$]${\sc i}   & 7&    +0.299 & 0.026 & EW   &35&  +0.365& 0.067  \\
$[$Sc/Fe$]${\sc ii}  & 7&  $-$0.088 & 0.034 & EW   &34&$-$0.011& 0.053  \\
$[$Ti/Fe$]${\sc i}   & 7&    +0.146 & 0.016 & EW   &  &        &        \\
$[$Ti/Fe$]${\sc ii}  & 7&    +0.168 & 0.023 & EW   &  &        &        \\
$[$V/Fe$]${\sc i}    & 7&  $-$0.129 & 0.026 & EW   & 9&$-$0.113& 0.051  \\
$[$Cr/Fe$]${\sc i}   & 7&  $-$0.211 & 0.029 & EW   &  &         &       \\
$[$Cr/Fe$]${\sc ii}  & 7&    +0.002 & 0.051 & EW   &  &        &        \\
$[$Mn/Fe$]${\sc i}   & 7&  $-$0.499 & 0.017 & EW   &  &        &        \\
$[$Fe/H$]${\sc i}    & 7&  $-$1.867 & 0.050 & EW   &35&$-$1.869& 0.093  \\
$[$Fe/H$]${\sc ii}   & 7&  $-$1.903 & 0.025 & EW   &  &        &        \\
$[$Co/Fe$]${\sc i}   & 7&  $-$0.007 & 0.023 & EW   &  &        &        \\
$[$Ni/Fe$]${\sc i}   & 7&  $-$0.106 & 0.023 & EW   &31&$-$0.116& 0.062  \\
$[$Cu/Fe$]${\sc i}   & 7&  $-$0.586 & 0.107 & synt &  &        &        \\
$[$Zn/Fe$]${\sc i}   & 7&    +0.013 & 0.048 & EW   &  &        &        \\  
$[$Y/Fe$]${\sc ii}   & 7&  $-$0.083 & 0.119 & synt &  &        &        \\
$[$Zr/Fe$]${\sc ii}  & 7&    +0.169 & 0.147 & synt &  &        &        \\
$[$Ba/Fe$]${\sc ii}  & 7&  $-$0.063 & 0.075 & EW   &  &        &        \\ 
$[$La/Fe$]${\sc ii}  & 7&    +0.107 & 0.064 & synt &  &        &        \\
$[$Nd/Fe$]${\sc ii}  & 7&    +0.450 & 0.059 & EW   &  &        &        \\ 
$[$Eu/Fe$]${\sc ii}  & 7&    +0.585 & 0.126 & synt &  &        &        \\ 
\hline
\end{tabular}
\label{meanabu}
\end{table}

\clearpage

\Online
\begin{table*}
\centering 
\setcounter{table}{1}
\caption{Information on the stars observed.}
\begin{tabular}{ccccccrcll}
\hline
\hline
 ID   &    RA       &  Dec         &  V     &   I     &  K	  &  RV    & err  & 2MASS ID \& Qflg & Note\\
      &             &              &        &         & (2MASS)& (km~s$^{-1}$) & (km~s$^{-1}$)&      & \\
\hline
\multicolumn{9}{c}{UVES} \\
2   & 217.3752792 & -5.9776094  & 14.776 & 13.295 & 11.476  &	-17.46 &  2.38 & 14293006-0558393, AAA & \\
5   & 217.3894665 & -6.0066020  & 14.967 & 13.593 & 11.833  &	-13.92 &  0.82 & 14293347-0600238, AAA & \\
7   & 217.4034627 & -5.9815808  & 15.196 & 13.901 & 12.491  &	-13.89 &  0.35 & 14293912-0558552, AAA & \\
9   & 217.4130141 & -5.9819918  & 15.404 & 14.125 & 12.491  &	-24.56 &  3.04 & 14293912-0558552, AAA & \\
10  & 217.4090060 & -6.0074072  & 15.421 & 14.165 & 12.486  &	-20.84 &  0.41 & 14293816-0600267, AAA & \\
12  & 217.4058084 & -5.9884596  & 15.437 & 14.212 & 12.648  &	-16.37 &  0.27 & 14293739-0559184, AAA & \\
13  & 217.4053172 & -5.9571719  & 15.505 & 14.258 & 12.658  &	-22.77 &  0.37 & 14293727-0557258, AAA & \\
\multicolumn{9}{c}{GIRAFFE, members} \\
100   & 217.4165390 &-5.9669299  & 17.056 &16.014 & 14.524  &	-12.88 &  0.89 & 14293996-0558009, AAA &     \\
103   & 217.3752167 &-5.9927176  & 17.069 &16.048 & 14.734  &	-16.59 &  0.43 & 14293005-0559337, AAA &     \\
113   & 217.4499290 &-6.0043198  & 17.152 &16.140 & 14.711  &	-13.11 &  0.99 & 14294799-0600155, AAA &     \\
125   & 217.3869503 &-5.9959868  & 17.324 &16.340 & 14.848  &	-18.63 &  0.85 & 14293286-0559456, AAB &     \\
132   & 217.3972697 &-6.0115462  & 17.357 &16.362 & 15.243  &	-14.41 &  0.88 & 14293533-0600416, AAC &     \\
135   & 217.3740604 &-6.0444890  & 17.399 &16.374 & 15.198  &	-15.82 &  0.87 & 14292978-0602398, AAC &     \\
142   & 217.3935980 &-5.9932600  & 17.397 &16.408 & 15.152  &	-10.14 &  0.80 & 14293445-0559357, AAC &     \\
146   & 217.4201662 &-6.0020515  & 17.425 &16.429 & 15.304  &	-14.29 &  1.05 & 14294084-0600074, AAC &     \\
151   & 217.4094269 &-6.0424978  & 17.525 &16.532 & 15.402  &	-23.04 &  0.46 & 14293826-0602328, AAC &     \\
152   & 217.3618344 &-5.9720939  & 17.526 &16.536 & 15.050  &	-17.01 &  0.81 & 14292683-0558195, ABC &     \\
163   & 217.3842792 &-5.9999402  & 17.701 &16.730 & 15.143  &	-20.97 &  0.96 & 14293223-0559597, BUU &     \\
167   & 217.3740827 &-5.9814592  & 17.768 &16.814 & 15.617  &	-18.95 &  0.91 & 14292982-0558532, BDD &     \\
169   & 217.4421960 &-5.9887997  & 17.795 &16.818 & 15.696  &	-14.70 &  0.71 & 14294615-0559194, ABD &     \\
173   & 217.4202268 &-5.9664150  & 17.840 &16.841 & 15.363  &	-19.01 &  0.81 & 14294082-0557591, BBC &     \\
182   & 217.3967426 &-5.9573177  & 17.891 &16.908 & 15.012  &	-16.95 &  0.88 & 14293521-0557263, ACU &     \\
189   & 217.4150008 &-6.0110400  & 17.923 &16.970 & 15.456  &	-10.40 &  0.69 & 14293959-0600399, BCU &     \\
194   & 217.3840477 &-6.0305388  & 17.961 &17.031 & 15.706  &	 -6.25 &  0.95 & 14293216-0601498, BCU &     \\
19    & 217.4173045 &-5.9842484  & 15.903 &14.722 & 13.253  &	-22.25 &  0.78 & 14294015-0559032, AAA &     \\
22    & 217.3961731 &-5.9896488  & 15.908 &14.760 & 13.295  &	-13.98 &  0.82 & 14293508-0559227, AAA &     \\
30    & 217.3945621 &-5.9777902  & 16.066 &14.969 & 13.676  &	-18.62 &  0.88 & 14293469-0558400, AAA & AGB \\
35    & 217.4141986 &-6.0033622  & 16.199 &15.069 & 13.545  &	-15.70 &  0.76 & 14293941-0600121, AAA &     \\
3     & 217.4188194 &-5.9526837  & 14.761 &13.329 & 11.491  &	-19.82 &  0.62 & 14294052-0557095, AAA &     \\
45    & 217.4092346 &-5.9703678  & 16.342 &15.276 &         &	-21.50 &  0.88 &    		       & AGB \\
46    & 217.4083105 &-5.9657353  & 16.332 &15.280 &         &	-12.46 &  1.16 &    		       & AGB \\
54    & 217.3991170 &-5.9643758  & 16.576 &15.477 & 14.045  &	-15.88 &  0.82 & 14293578-0557517  AAA &     \\
58    & 217.4053660 &-5.9919049  & 16.616 &15.533 & 14.104  &	-14.81 &  0.91 & 14293728-0559309  AAA &     \\
60    & 217.3846260 &-5.9826466  & 16.622 &15.555 & 14.030  &	-12.81 &  0.94 & 14293230-0558575  AAA &     \\
62    & 217.4241485 &-5.9672889  & 16.663 &15.576 & 14.178  &	-14.73 &  0.82 & 14294180-0558022  AAA &     \\
64    & 217.4016268 &-5.9749167  & 16.629 &15.611 & 14.344  &	-15.16 &  0.71 & 14293640-0558300  AAB & AGB \\
70    & 217.4114357 &-5.9659564  & 16.773 &15.684 & 14.438  &	-15.54 &  0.89 & 14293874-0557574  AAA &     \\
72    & 217.4025520 &-5.9682492  & 16.738 &15.718 & 14.633  &	-13.70 &  0.76 & 14293661-0558056  AAB & AGB?\\
78    & 217.4060006 &-5.9783599  & 16.803 &15.770 &         &	 -8.19 &  0.78 &    		       &     \\
82    & 217.3991885 &-5.9850201  & 16.736 &15.801 & 14.433  &	-15.54 &  0.66 & 14293580-0559061  AAA & AGB \\
87    & 217.3906596 &-5.9562227  & 16.909 &15.862 & 14.388  &	-14.52 &  1.01 & 14293375-0557223  AAA &     \\
93    & 217.4148681 &-5.9978062  & 16.972 &15.927 & 14.697  &	-20.58 &  0.82 & 14293956-0559521  AAB &     \\
97    & 217.3781654 &-6.0014762  & 16.907 &15.960 & 14.969  &	-12.27 &  0.95 & 14293075-0600053  AAB & AGB \\
\multicolumn{9}{c}{GIRAFFE, non members/unknown} \\
11    & 217.3394083 &-6.0043047  & 15.407 &14.200 &            &	 56.94 &  0.35 &                                          &     \\
10026 & 217.5510953 &-5.9215522  & 17.666 &16.675 &         &	       &       &    		       &     \\
\hline
\end{tabular}
\label{tabM}
\end{table*}

\end{document}